\documentclass{article}
\usepackage{bm}
\usepackage{amssymb,amsfonts}

\def\div{\nabla \cdot}

\def\W{{\Omega}}
\def\vf{{\varphi}}
\def\op{{\varphi}}
\def\k{{\kappa}}
\def\l{{\lambda}}

\def\b{{\beta}}
\def\g{{\gamma}}

\def\e{{\varepsilon}}

\def\bn{{\bi n}}

\def\bv{{\bi v}}
\def\bw{{\bi w}}

\def\bzero{{\bi 0}}

\def\R{{\mathbb R}}

\def\bea{\begin{eqnarray*}}
\def\eea{\end{eqnarray*}}
\def\rmd{d}
\def\rme{e}
\def\fl{}
\def\boldsymbol{\bm}

\newtheorem{theor}{Theorem}[section]
\newtheorem{lem}{Lemma}[section]
\newtheorem{prop}{Proposition}[section]

\def\qed{\hfill $\square$}

\def\W{{\Omega}}
\def\vf{{\varphi}}
\def\k{{\kappa}}
\def\l{{\lambda}}

\def\b{{\beta}}
\def\g{{\gamma}}

\def\e{{\varepsilon}}

\def\bD{{\bf D}}

\def\bJ{{\bf J}}
\def\bT{{\bf T}}
\def\bL{{\bf L}}

\def\bQ{{\bf Q}}

\def\bd{{\bf d}}

\def\bh{{\bf h}}
\def\bn{{\bf n}}

\def\bv{{\bf v}}
\def\bw{{\bf w}}
\def\bzero{{\bf 0}}

\def\R{{\mathbb R}}

\begin{document}
\author{A. Berti\footnote{Faculty of Engineering, University e-campus,  22060 Novedrate (CO), Italy,
e-mail: alessia.berti@ing.unibs.it}, V. Berti\footnote{University of Bologna, Department of Mathematics, 40126 Bologna, Italy, e-mail: berti@dm.unibo.it}, D. Grandi\footnote{University of Bologna, Department of Mathematics, 40126 Bologna, Italy, e-mail: grandi@dm.unibo.it}
}
\title{Well posedness of an isothermal diffusive model\\ for binary mixtures of incompressible fluids}
\date{}
\maketitle
\begin{abstract}
\noindent
We consider a model describing the behavior of a mixture of two incompressible fluids with the same density in isothermal conditions. The model consists of three balance equations: continuity equation,  Navier-Stokes equation for the mean velocity of the mixture, and diffusion equation (Cahn-Hilliard equation).
We assume that the chemical potential depends upon the velocity of the mixture in such a way that an increase of the velocity improves the miscibility of the mixture.  We examine the thermodynamic consistence of the model which leads to the introduction of an additional constitutive force in the motion equation.
Then, we prove existence and uniqueness of the solution of the resulting differential problem.
\end{abstract}

\bigskip


\bigskip


\date{}

\maketitle

\section{Introduction}

We consider a model describing the isothermal motion of a mixture of two incompressible fluids following the \emph{diffusional approach} to binary mixtures. This goes back to \emph{model H} in the classification by Hohenberg and Halperin \cite{HH}, which consists of a Cahn-Hilliard diffusion model coupled with a fluid motion. This kind of approach is extensively discussed by Lowengrub and Truskinovsky \cite{Truski}. Basically, a binary mixture is described in terms of a macroscopic velocity field representing the mean velocity at each spatial point, the total-density field of the mixture, and an order field which describes the actual composition of the mixture at each point. The model consists of three balance equations: continuity equation, (total) momentum balance (Navier-Stokes equation), and diffusion equation (Cahn-Hilliard equation). The diffusional model can be considered as an approximation of the classical theory of mixture (based on two continuity equations and two momentum balance equations, one for each component) when the momenta and the kinetic energies of the relative motion can be neglected. \\
The  coupling of the motion equation with the Cahn-Hilliard equation has a trivial part due to the presence of the material derivative (rather than the partial time derivative) of the order parameter in the Cahn-Hilliard equation and a less trivial coupling arising from the dependence of stress tensor upon the gradient of the order parameter. The presence of such a coupling has been derived by Gurtin et al. on the basis of classical continuum mechanics arguments \cite{Gurtin}.\\
In this paper we discuss a variant of this model in which a dependence of the chemical potential upon the velocity of the mixture is introduced. 
We add the velocity-dependent term in the local part of the chemical potential, that is the part independent from the gradients of the order parameter. 
The effect of the velocity can be assimilated to an increase of the temperature (which is a fixed parameter in the isothermal model we are considering), that is it reduces the miscibility gap.

In section \ref{sec:model} we review the classical analysis of the thermodynamic consistence which displays the need of an additional constitutive force term in the motion equation.\\
The subsequent sections are devoted to the proof of existence and uniqueness of the solution of the resulting differential problem.
Our mathematical study concentrates on the \emph{fully incompressible} situation, that is the case of a binary mixture of two incompressible fluids which also have the same density. Clearly, this is an exceptional case from an empirical point of view, but could be an acceptable approximation for a broader class of real situations.

The coupling of the Navier-Stokes equations with the Cahn-Hilliard equation has been extensively studied in the literature. Among the first results concerning existence, uniqueness and asymptotic behaviour of the solutions, we recall the papers \cite{boyer} and \cite{Staro}. More recently, Abels \cite{Abels} proves well-posedness and examines long-time behaviour of the Cahn-Hilliard-Navier-Stokes system involving a class of singular free energies, (including the logarithmic free energy) which guarantees the boundedness of the order parameter. Further results about the asymptotic behaviour of the solutions and the existence of global and exponential attractors for the coupled system are shown in \cite{GG}.

The main result of our paper is the proof of existence and uniqueness of the solutions of Cahn-Hilliard-Navier-Stokes equations where a new non-linear term is present due to the velocity-dependence of the chemical potential.
Moreover, we add a viscous term in the definition of the chemical potential (\cite{Gurt}) which turns out to be crucial for our purpose. 
The functional formulation of our problem is given in section \ref{sec:functional}.
In section \ref{sec:approx}, with the same technique used in \cite{Gal}, we introduce a family of approximating problems, by adding to the Cahn-Hilliard equation a perturbative term proportional to the time derivative of the chemical potential.
By means of a fixed point argument, we establish the existence of solutions of the approximated problems. Finally, in section \ref{sec:original} we prove well-posedness of the original problem letting the perturbative term tend to zero.

\section{Model equations and thermodynamical consistence} \label{sec:model}

Let us consider a mixture of two partially miscible fluids; we will use a binary  index $i=1,2$ to make reference to each of them. Every spatial volume element $\rmd V$ will in general contain a mass portion $\rmd m_i$ of the $i$-th fluid; we indicate with $\rho_i$ the apparent density of each fluid:
$$
\rho_i=\frac{\rmd m_i}{\rmd V}.
$$
The adjective `apparent' is used to emphasize that we are considering the ratio of each mass fraction over the total volume element $\rmd V$, rather than over its own fractional volume $\rmd V_i$. Each component, which can be compressible or incompressible, is characterized by its own density $\rho_i^0=\rmd m_i/\rmd V_i$ in standard conditions of temperature and pressure. 
Of course, the total density is the sum $\rho=\rho_1+\rho_2$. We also define an \emph{order parameter} measuring the degree of phase separation as 
$$
\vf=\frac{\rmd m_1-\rmd m_2}{\rmd m_1+\rmd m_2}=\frac{\rho_1-\rho_2}{\rho}\in[-1,1].
$$
As we are considering mutually non-transforming chemical species, the first general balance laws we have to impose are the mass conservations of each component; so, if $\bv_i$ is the velocity of the $i$-component, we demand
\begin{equation}\label{eqn:cont-i}
\frac{\partial\rho_i}{\partial t}+\nabla\cdot(\rho_i\bv_i)=0, \quad i=1,2.
\end{equation}
By defining the \emph{mean velocity} of the mixture as 
$$
\bv=\rho^{-1}(\rho_1\bv_1+\rho_2\bv_2),
$$
(so that $\rho\bv$ is the total momentum density), a global mass continuity law follows
\begin{equation}\label{eqn:cont}
\frac{\partial\rho}{\partial t}+\nabla\cdot(\rho\bv)=0.
\end{equation}
The model we are going to study describes an incompressibile mixture of two fluids, so the total density $\rho=\rho_1+\rho_2$ is constant and the continuity equation (\ref{eqn:cont}) reduces to
 $$\nabla\cdot\bv=0.$$ This means that each fluid component is an incompressible fluid and also that each component has the same constant density $\rho$:
$$
 \rmd m_i=\rho \rmd V_i.
$$

From equations (\ref{eqn:cont-i}) it also follows, after a little calculation, the following equation for $\op$:
$$
\rho\left(\frac{\partial \op}{\partial t}+\bv\cdot\nabla \op\right)=\nabla\cdot\left[\frac{2\rho_1\rho_2}{\rho}(\bv_2-\bv_1)\right].
$$
We will use the usual notation for the material time derivative related to the \emph{mean} velocity field $\bv$ (\emph{barycentric material derivative}):
$$
\dot{f}\equiv \frac{\partial f}{\partial t}+\bv\cdot\nabla f,
$$
for every spatial field $f(x,t)$. Therefore the equation for $\op$ assumes the form 
\begin{equation}\label{eqn:evol-op}
\rho\dot \op=\div\bJ.
\end{equation}
This is the usual equation for a conserved quantity with respect the gross  motion of the fluid defined by the mean velocity $\bv$.\\
We  introduce now a basic physical hypothesis which characterizes the diffusional approach to binary fluids (\cite{Gurtin, Truski}). Accordingly, we will describe the dynamics by using a balance law for the \emph{total momentum density} $\rho\bv$ of the mixture, while the effects of the relative motion will be described only through the balance law (\ref{eqn:evol-op}) for the scalar \emph{order field} $\op$, not retaining the motion equation for the relative momentum $\rho_2\bv_2-\rho_1\bv_1$. In other words, the fundamental fields of the model will be $\rho, \bv, \op$ rather than $\rho_1,\rho_2, \bv_1, \bv_2$, and the current $\bJ$ will be given a constitutive law in terms of $\rho, \bv, \op$ (and their gradients). Physically, this amounts to neglect the kinetic energies and the momenta of the constituents relative to the mean motion, only retaining the information of the relative mass flux. The resulting model can be considered as the model of a single fluid with an internal  defined by the conserved field $\op$.

 So in this paper we are going to consider a model characterized by the three balance equations:
\begin{eqnarray}\label{eqn:balances}
 \left\{\begin{array}{ll}
 \nabla\cdot\bv=0,\\
 \rho\dot{\bv}=\nabla\cdot\bT+\rho\bd+\rho\bf f,\\
 \rho\dot \op=\nabla\cdot \bJ.
 \end{array}\right.
\end{eqnarray} 
Here $\bT$ is the stress tensor, $\bd$ a constitutive body force and  $\bf f$ a possible external body force. The (unusual) constitutive force $\bd$ is required if we want to construct a thermodynamically consistent model in which a dependence of $\bJ$ on $\bv$ is admitted, as we are going to show.

We begin considering the diffusion equation. We will consider a Cahn-Hilliard similar model, in which the current is expressed as
\begin{equation}\label{J}
\bJ=\gamma\nabla\mu,
\end{equation}
where $\mu$ is the chemical potential. In the classical Cahn-Hilliard model  (\cite{Cahn, CH}) $\mu$ is a non local function of the order field $\op$ which takes the form
$$
 \mu=\mu_{\rm loc}(\op)-\nabla\cdot(\kappa\bh),
$$
where $\bh=\nabla\op$. Here we want to consider a generalization which possibly includes a dependence on velocity $\bv$. Also, for future utility in the mathematical study, we will add a dissipative contribution proportional to $\dot\op$. So we assume
$$
\mu=\mu_{\rm loc}(\bv,\op)-\kappa_1(\op)\nabla\cdot[\,\kappa_2(\op)\bh\,]+\beta\dot\op.
$$
 
We associate a suitable balance of powers to the diffusion equation, which is needed to write the first law of thermodynamics later on. This is obtained by multiplying the diffusion equation by the chemical potential $\mu$ 
\begin{eqnarray}\label{eqn:pow-bal-op}
\rho\dot \op\mu=\mu\nabla\cdot\bJ.
\end{eqnarray}
The central issue is to recognize in this equality an \emph{internal} and an \emph{external} power. The choice will be influential in satisfying the second law of thermodynamics. In particular, as there are non local (gradient) contributions in $\mu$, it would create difficulties to refer completely the term $\rho\dot \op\mu$ to the internal power. 

We will rewrite equation (\ref{eqn:pow-bal-op}) in the form
$$
\rho\dot \op\mu_{\rm  loc}+\beta\rho\dot \op^2+\kappa_2\bh\cdot\nabla(\rho\kappa_1\dot \op)+\bJ\cdot\nabla\mu=\nabla\cdot(\mu\bJ+\rho\kappa_1\kappa_2\dot \op\bh)
$$
and we ascribe the left hand side to the internal power $\mathcal P^i_\op$.\footnote{There is an alternate approach to the issue of the energy balance associated to the diffusion equation due to Gurtin. According to this author, the diffusion equation is associated to the internal power $\rho\dot \op\mu+\bJ\cdot\nabla\mu$, which could be considered as a definition of the chemical potential, which is treated as an independent field. Moreover the diffusion equation is accompained with an independent balance equation (microforce balance) which also brings a contribution to the total amount of power.} Letting $\bL=\nabla\bv$ (in components: $L_{ij}=\partial v_i/\partial x_j$), we observe that
$$
\nabla\dot \op=\dot\bh+\bL^T\bh.
$$
We use this identity and (\ref{J}) to write the internal power in the more suited form
$$
\mathcal P^i_\op=[\rho\mu_{\rm  loc}+\kappa_2\bh\cdot\nabla(\rho\kappa_1)]\dot \vf+\beta\rho\dot \vf^2+\rho\kappa_1\kappa_2\bh\cdot(\dot\bh+\bL^T\bh)+\gamma|\nabla\mu|^2.
$$

Next we consider the momentum balance equation. 
The stress tensor and the constitutive force are given by
$$
\bT=\hat\bT(\bD,\op,\bh),\qquad \bd=\hat\bd(\bv,\op,\dot \op),
$$
where $\bD:=\frac{1}{2}(\bL+\bL^T)$ is the symmetric part of the velocity gradient.\\
The balance of powers is obtained by multiplying both members of (\ref{eqn:balances})$_2$ by $\bv$:
$$
\frac{1}{2}\rho(\bv^2)^\cdot=[\nabla\cdot(\bT\bv)+ \rho\bf f\cdot\bv]-[\bT:\nabla\bv-\rho\bd\cdot\bv].
$$
The term $\bd\cdot\bv$ from the constitutive body force will contribute to the internal power. In particular, the internal mechanical power $\mathcal P^i_{\rm m}$ is defined by
$$
\mathcal P^i_{\rm m}=\bT:\bD-\rho\bd\cdot\bv.
$$
This identification is based on the assumption that stress tensor is a function depending only upon the first gradients of the field, namely $\bT=\hat\bT(\bD,\op,\bh)$; otherwise, as it happens for $\mu$, which is dependent upon $\nabla\bh$, it would be appropriate to refer part of the contribution $\bT:\bD$ to the external power. 

As we are considering an isothermal model, we use the \emph{dissipation inequality}
$$
\rho\dot\psi-\mathcal P^i_{\op}-\mathcal P^i_{\rm m}\leq0,
$$
where $\psi$ the free energy, as the proper version of the second law of thermodynamics.\\
If  $\psi=\hat\psi(\chi)$, where $\chi$ is the list of the variables  $\bv,\op,\dot\op$ and all their gradients appearing in the constitutive equations, the dissipative inequality is written as
\begin{eqnarray*}
&&\fl\rho\sum_{\chi_i\neq\op,\bh}\psi_{\chi_i}\dot{\chi_i}\,+[\rho\psi_\op-\rho\mu_{\rm loc}-\kappa_2\bh\cdot\nabla(\rho\kappa_1)]\,\dot \op-\beta\rho\,\dot \op^2+[\rho\psi_{\bh}-\rho\kappa_1\kappa_2\bh]\cdot\dot\bh\\
&&-[\bT+\rho\kappa_1\kappa_2\bh\otimes\bh]:\bD+\rho\bd\cdot\bv-\gamma|\nabla\mu|^2\leq 0,
\end{eqnarray*}
having used $\bh\cdot(\bL^T\bh)=\bD:(\bh\otimes\bh)$. It easily follows that $\psi$ does not depend upon any of the variables $\chi_i\neq\op,\bh$, that is
\begin{equation}\label{eqn:psi}
\psi=\hat\psi(\op,\bh).
\end{equation}

For any given $\op,\bh$, it is possible to find processes with $\dot\op=0,\bv={\bf 0},\bD={\bf 0},\nabla\mu={\bf 0}$ but otherwise with $\dot\bh$ arbitrary.\footnote{In fact $\nabla\mu=\nabla(\mu_{\rm loc}+\beta\dot\op)+\nabla[\kappa_2\nabla\cdot(\kappa_1\bh)]$, so one can make $\nabla\mu=0$ by suitably choosing $\nabla\nabla\bh$ for the given set of conditions.} This implies
\begin{equation}\label{eqn:psi2}
 \psi_\bh=\kappa_1\kappa_2\bh.
\end{equation}
By choosing appropriately $\dot\bh$ we can make $\nabla\mu={\bf 0}$ with $\op,\dot\op,\bh,\bv,\bD$ arbitrary, so the following inequality holds
\begin{eqnarray}
&&\fl[\rho\psi_\op-\rho\mu_{\rm loc}-\kappa_2\bh\cdot\nabla(\rho\kappa_1)]\,\dot \op-\beta\rho\dot\op^2-[\hat\bT(\bD,\op,\bh)+\rho\kappa_1\kappa_2\bh\otimes\bh]:\bD\nonumber\\
&&+\rho\hat\bd(\bv,\op,\dot \op)\cdot\bv\leq 0.
 \label{eqn:red-diseq}
 \end{eqnarray}
In the same way, we can make $\nabla\mu\neq0$ and $\dot\op=0,\bv={\bf 0},\bD={\bf 0}$, so $\gamma>0$.\\
Inequality (\ref{eqn:red-diseq}) implies that $\beta>0$ (considering processes with $\bv={\bf 0}$, $\bD={\bf 0}$ and  $\dot\op$ arbitrary); letting $\dot\op=0$ and $\bv={\bf 0}$ we have to impose
$$
 [\hat\bT(\bD,\op,\bh)+\rho\kappa_1\kappa_2\bh\otimes\bh]:\bD\geq 0.
$$

Because of the incompressibility, on one hand the pressure (that is the trace part of $\bT$) is not a constitutively determined quantity (it is kinematically determined), on the other hand the trace part of $\bD$ identically vanishes
$$
 Tr(\bD)=\nabla\cdot \bv=0.
$$
So,  putting for brevity, 
$$
\bQ\equiv\bT+\rho\kappa_1\kappa_2\bh\otimes\bh,
$$
the inequality $\bQ:\bD\geq 0$ is equivalent to $\tilde\bQ:\tilde\bD\geq 0$
where the tilde on a tensor indicates its deviatoric part: $\tilde\bD=\bD-\frac{1}{3}Tr(\bD){\bf 1}$ and similarly for $\bQ$. 
So we assume $\tilde\bQ=2\nu\tilde\bD$ with $\nu=\hat\nu(\bD,\op,\bh)>0$, that is
\begin{equation}\label{eqn:T-inc}
\bT+\rho\kappa_1\kappa_2\bh\otimes\bh=2\nu\tilde\bD-p{\bf 1}.
\end{equation}
where $p=-\frac{1}{3}Tr(\bQ)$ is the indetermined component of the pressure.\\
We are now left with the reduced inequality (for $\bD=0$)
\begin{equation}
 \label{eqn:red-diseq2}
 \fl[\rho\psi_\op(\op,\bh)-\rho\mu_{\rm loc}(\op,\bv)-\kappa_2\bh\cdot\nabla(\rho\kappa_1)]\,\dot \op-\beta\rho\dot\op^2+\rho\hat\bd(\bv,\op,\dot \op)\cdot\bv \leq 0.
\end{equation}

From that we see that, if there is a non trivial dependence of $\mu_{\rm loc}$ on $\bv$, the presence of the constitutive force $\bd$ is necessary. The following particular choices are made to satisfy (\ref{eqn:red-diseq2}):
\begin{eqnarray}
 \bd=\boldsymbol{\delta}(\bv,\op)\dot\op,\label{eqn:d}\\
\kappa_1={\rm constant},\quad\psi_{\op\bh}=0,\\
\mu_{\rm loc}(\rho,\op,\bv)-\boldsymbol{\delta}\cdot\bv=\psi_\op(\op)\label{eqn: muloc}
\end{eqnarray}
Conditions (\ref{eqn:psi}),  (\ref{eqn:psi2}), (\ref{eqn:T-inc}),(\ref{eqn:d})-(\ref{eqn: muloc}) with $\beta>0,\gamma>0$, ensure that the dissipation inequality is satisfied.

The specific feature of the model we are going to study (compared with those by Gurtin and Truskinovsky) is a velocity dependence of the chemical potential. We adopt the usual form of the free energy function $\hat\psi(\op,\bh)$ used in the Cahn-Hilliard model of diffusion:
$$
\psi=\frac{\kappa}{2}|\nabla\op|^2+u\cdot G(\op)+H(\op),\quad G(\op)\equiv\frac{1}{2} \op^2,\quad H(\op)\equiv \frac{1}{4}\op^4,
$$
where $u\in[-1,+\infty[$ is a temperature-dependent parameter, typically 
$$
u=\frac{\theta-\theta_c}{\theta_c}.
$$
The $\vf$-dependent part of $\psi$ is such that for $u\geq0$ it has a unique minimum at $\op=0$, while for $-1\leq u<0$ has two minima $\pm \bar \op$ with $\bar \op\in ]0,1]$. It is known (\cite{CH}) that the unique minimum in the potential corresponds to the situation without \emph{miscibility gap}, while in the regime with two minima there is a miscibility gap.\\
In this paper we assume the following form for the local part of the chemical potential
$$
\hat\mu_{\rm loc}(\op,\bv)=\hat\psi_\op(\op)+\lambda\bv^2 G'(\op).
$$
We remark that the effect of velocity can be assimilated with an increase of temperature, that is a restriction of the miscibility gap.  
So $\boldsymbol{\delta}\cdot\bv=\lambda\bv^2\, G'(\op)$ and we obtain the constitutive force
$$\bd=\lambda\bv\, G'(\op)\dot \op=\lambda\bv\, \dot G(\op).
$$

We sum up the system of equation in that case, putting everywhere $\rho=1$ and taking $\gamma,\nu,\kappa=\kappa_1\kappa_2>0$ constant:
\begin{eqnarray}
\label{cont}
&&\div \bv = 0,\\
\label{dot-v}
&&\dot\bv = -\nabla p + \nu \Delta\bv - \kappa \div (\nabla\op \otimes \nabla\op) + \l \op\dot \op\bv+\mathbf f,
\\
\label{dot-phi}
&&\dot \op =\gamma \Delta\mu,
\end{eqnarray}
where
\begin{eqnarray}\label{mu}
	\mu = -\kappa \Delta \op+ \op^3 + (u+\lambda \bv^2)\op+\beta\dot \op.
\end{eqnarray}

\section{Notation and functional settings}\label{sec:functional}

In order to obtain a precise formulation of the problem, we introduce here some notation and recall the main inequalities used in the sequel. We assume that the domain $\W$
 occupied by the system is a bounded
subset of $\mathbb R^2$, with smooth boundary $\partial \W$.

For each $p\geq 1$ and $s\in\R$, we denote by $L^p(\W)$ and $H^s(\W)$ the Lebesgue and Sobolev spaces of real valued or vector valued functions, according to the context. Let $\|\cdot\|_p$ and $\|\cdot\|_{H^s}$ be the standard norms of $L^p(\W)$ and $H^s(\W)$, respectively. In particular $\|\cdot\|$ stands for the $L^2(\W)$-norm.
The space $H^1_0(\W)$ is the closure of $C^\infty$ functions with compact support with respect to the norm $\|\cdot\|_{H^1}$. Moreover, $H^{1}(\W)'$ is the dual space of $H^1(\W)$ endowed with the standard norm
\begin{equation}\label{-1norm}
	\|w\|_{(H^{1})'} = \sup\{|\langle w,u\rangle| : u\in H^1(\W), \|u\|_{H^1}\leq 1 \},
\end{equation}
where $\langle \cdot, \cdot \rangle$ denotes the pairing.

We define 
\bea
\widehat H^1(\W) &=& \left\{ w \in H^1(\W) : \nabla w\cdot \bn|_{\partial\W}  =0 \right\},
\\
\widehat H^2(\W) &=& \left\{ w \in H^2(\W) : \nabla w\cdot \bn|_{\partial\W}  =0 \right\}.
\eea
For vector valued functions, we introduce the functional spaces used in the framework of Navier-Stokes equations (\cite{Tem_NS})
\bea
H^1_{\rm div}(\W) &=& \left\{ \bw \in H^1_0(\W) : \div\bw =0\right\},\\
L^2_{\rm div}(\W) &=& \left\{ \bw \in L^2(\W) : \div\bw =0, \bv\cdot\bn|_{\partial\W}=0\right\}.
\eea
Finally, for any $T>0$ we define
\bea
X_{\vf} &=& L^2(0,T; H^3(\W) \cap \widehat H^2(\W)) \cap H^1(0,T; H^1(\W)) \cap H^2(0,T;H^{1}(\W)')
\\
X_{\bv} &=& L^2(0,T; H^2(\W) \cap H^1_{\rm div}(\W)) \cap H^1(0,T; L^2_{\rm div}(\W))
\\
X_{\mu} &=& L^2(0,T; \widehat H^2(\W)) \\
X_T&=&X_{\vf}\times X_{\bv}\times X_{\mu}
\eea
endowed with their usual norms $\|\cdot\|_{X_{\vf}}$, $\|\cdot\|_{X_{\bv}}$, $\|\cdot\|_{X_{\mu}}$ and 
$$
\|(\vf,\bv,\mu)\|_{X_{T}}^2 = \|\vf\|_{X_{\vf}}^2 + \|\bv\|_{X_{\bv}}^2 + \|\mu\|_{X_{\mu}}^2.
$$

Here and henceforth we denote by $C$ any positive constant depending only on the domain $\W$ which is allowed to vary even in the same formula. Further dependencies will be specified.

Since $\Omega \in \mathbb{R}^2$ the Sobolev embedding theorem implies (\cite{A})
\begin{equation}
\label{Sob1}
\|w\|_p\leq C \|w\|_{H^1},\qquad 2\leq p< \infty,\quad\ w\in H^1(\W)
\end{equation}
and the following interpolation inequalities hold as a consequence of the Ga\-gliar\-do-Nirenberg inequality \cite{gagliardo,nirenberg}:
\begin{eqnarray}\label{L4}
	\|w\|_4^2 &\leq& C \|w\| \| w\|_{H^1},
\\
\label{L6}
	\|w\|_6^2 &\leq& C \|w\|^{4/3} \| w\|^{2/3}_{H^1}.
\end{eqnarray}

If $w\in H^1_0(\W)$ or $w\in H^1(\W)$ and $\int_\W w \rmd x=0$, the Poincar\'e inequality provides (\cite{E})
$$\|w\|\leq C\|\nabla w\|\,.
$$
From the Agmon inequality (\cite[p.52]{Tem}), we deduce that
\begin{equation}
\label{infty}
\|w\|_\infty \leq C \|w\|_{H^2},\qquad  w\in H^2(\W).
\end{equation}
Furthermore, for every $v\in H^1(\W)$, $u, w\in H^2(\W)$ the following interpolation inequalities
\begin{eqnarray}\label{H1H2}
    \|v w\|_{H^1} &\leq& C \|v\|_{H^1} \|w\|_{H^2}
    \\
    \label{H2H2}
    \|u w\|_{H^2} &\leq& C \|u\|_{H^2} \|w\|_{H^2}
\end{eqnarray}
hold as a straightforward consequence of (\ref{Sob1}) and (\ref{infty}).

In addition, if $w\in H^2(\W) \cap H^1_0(\W)$ or $w \in \widehat H^2(\W)$, then (\cite[Thm. 5.1, pag. 149]{LM})
$$	\| w\|_{H^2} \leq C(\|w\| + \|\Delta w\|).
$$
As a consequence, for every $w\in H^3(\W)\cap \widehat H^2(\W)$, we have
\begin{equation}\label{H3}
	\| w\|_{H^3} \leq C(\|w\|_{H^1}+\|\Delta w\|_{H^1}) \leq C(\|w\|+\|\nabla\Delta w\|).
\end{equation}

For vector-valued functions we define the orthogonal projector $\mathcal{P}:L^2(\W)\to L^2_{\rm div}(\W)$ and the operator
$A$ defined as
$$
A\bw=-\mathcal{P}\Delta\bw, \qquad \bw\in H^2(\W)\cap H^1_{\rm div}(\W).
$$
It is worth noting that for any $\bw\in H^2(\W)\cap H^1_{\rm div}(\W)$ the following estimate holds:
$$
\|\bw\|_{H^2} \leq C(\|\bw\| + \|A\bw\|).
$$

For later use, we will also need the following result, whose proof is given in \cite[Thm 4, pag. 288]{E}.
\begin{theor}\label{thm-evans}
Suppose $w \in L^2(0,T;H^{m+2}(\W))$ and $w_t \in L^2(0,T;H^m(\W))$ where $m$ is a nonnegative integer. Then
$w \in C([0,T];H^{m+1}(\W))$ and
$$
\max_{[0,T]}\|w(t)\|_{H^{m+1}} \leq C(\|w\|_{L^2(0,T;H^{m+2}(\W))} + \|w_t\|_{L^2(0,T;H^m(\W))}),
$$ 
the constant C depending only on $T, \W$, and $m$. 
\end{theor}

Finally, for reader's convenience, we recall Young's inequality.
Let $1<p,q<\infty$, such that $\frac{1}{p} + \frac{1}{q} =1$. Then, 
\begin{equation}\label{young}
	ab \leq \eta a^p + C(\eta) b^q, \qquad (a,b>0, \eta>0).
\end{equation}

The functional formulation of problem (\ref{cont})-(\ref{mu}) is the following:

\noindent
{\bf Problem $(P)$.}
To find a triplet $(\vf,\bv,\mu) \in X_{T}$ such that
\begin{eqnarray}
\label{wk1}
&&\fl\beta \vf_t  - \k \Delta \vf  -\mu + \beta \bv \cdot \nabla \vf + \vf^3 +u\vf+ \lambda \vf \bv^2 =0
	\\
\label{wk2}
&&\fl\bv_t + \nu A \bv +\mathcal{P}\left[ \k \div (\nabla\vf \otimes \nabla\vf) - \lambda \vf (\vf_t + \bv\cdot \nabla \vf) \bv + (\nabla \bv) \bv\right]
=\mathcal{P}{\bf f}
	\\
\label{wk3}
&& \fl-\g \Delta\mu +\vf_t + \bv \cdot \nabla \vf =0
	\end{eqnarray}
a.e. in $\W\times(0,T)$, and
$$
\vf(x,0)=\vf_0(x), \qquad \bv(x,0)=\bv_0(x), \qquad a.e.\ x\in \W.
$$
Notice that equation (\ref{wk1}) which provides the definition of the chemical potential, is interpreted as a parabolic equation governing the evolution of $\vf$, whereas equation (\ref{wk3}) is an elliptic equation for the unknown $\mu$.
Accordingly, the condition $\b>0$ will be crucial in the following analysis to prove well-posedness of the system.

Existence of solutions to problem $(P)$ is proved by introducing a suitable family $(P_\e)$ of approximating problems, where $\e$ is a small parameter such that $0<\e<1$. In Section \ref{sec:approx} we prove existence of  solutions to $(P_\e)$. Then, by letting $\e\to 0$, we obtain the existence result for the solutions of problem $(P)$.

\section{Approximating problem}\label{sec:approx}
We construct a family of approximating problems of $(P)$, by adding the term
$\e\mu_t$ to equation (\ref{wk3}).
Accordingly, we introduce the functional space
$$
X_{T}^\e=X_\vf\times X_\bv\times X_{\mu}^\e,
$$
where
$$
X_{\mu}^\e = L^2(0,T; \widehat H^2(\W)) \cap H^1(0,T; L^2(\W)).
$$
\medskip

\noindent
{\bf Problem $(P_\e)$}. To find a triplet $(\vf^\e,\bv^\e,\mu^\e) \in X_{T}^\e$ such that
\begin{eqnarray}
\label{weak1}
&&\fl\beta \vf_t^\e  - \k \Delta \vf^\e  - \mu^\e + \beta \bv^\e \cdot \nabla \vf^\e + (\vf^\e)^3 +u\vf^\e + \lambda \vf^\e (\bv^\e)^2 = 0
	\\
&&	\fl\bv_t^\e + \nu A \bv^\e +\mathcal{P}[ \k \div (\nabla\vf^\e \otimes \nabla\vf^\e) - \lambda \vf^\e (\vf_t^\e + \bv^\e\cdot \nabla \vf^\e) \bv^\e + (\nabla \bv^\e) \bv^\e]
	=\mathcal{P}{\bf f}
	\label{weak2}
	\\
\label{weak3}
&&\fl\e \mu_t^\e - \g \Delta\mu^\e +\vf_t^\e + \bv^\e \cdot \nabla \vf^\e = 0 
\end{eqnarray}
a.e. in $\W\times(0,T)$, and
\begin{equation}\label{in-data}
\vf(x,0)=\vf_0(x), \  \bv(x,0)=\bv_0(x), \ \mu(x,0)=\mu_0(x),\  a.e.\ x\in \W,
\end{equation}
where $\mu_0$ is the solution to equation 
$$
-\beta\gamma\Delta\mu_0+\mu_0=-\k\Delta\vf_0+\vf^3_0+u\vf_0 + \lambda \vf_0 \bv^2_0. 
$$

From the standard theory of linear parabolic equations (see {\it e.g.} \cite{E,LM, Tem_NS}), we deduce the following auxiliary result.

\begin{lem}[Existence of solutions of problem $(P_L)$]\label{PL}
Let $\Phi \in L^2(0,T; H^1(\W))$ 

\noindent $\cap H^1(0,T;  H^{1}(\W)')$, $\Upsilon \in L^2(0,T; L^2_{\rm div}(\W))$, $\Lambda \in L^2(0,T; L^2(\W))$, $\vf_0 \in \widehat H^2(\W)$, $\bv_0 \in H^1_{\rm div}(\W)$, $\mu_0\in \widehat H^1(\W)$. Then, there exists a unique solution $(\vf,\bv,\mu) \in X_T^\e$ of the linear problem $(P_L)$
\begin{eqnarray}\label{vf-lin}
	\beta \vf_t -\k \Delta \vf &=& \Phi
	\\
	\label{v-lin}
	\bv_t + \nu A\bv &=& \Upsilon
	\\
	\label{mu-lin}
	\e \mu_t - \g \Delta\mu &=& \Lambda
\end{eqnarray}
with the initial conditions (\ref{in-data}).
In particular, $\vf \in C([0,T]; \widehat H^2(\W))$, $\bv  \in C([0,T]; H^1_{\rm div}(\W))$, $\mu \in C([0,T]; \widehat H^1(\W))$.
\end{lem}

\begin{lem}\label{lemma-lambda}
Let $(\psi, \bw, \zeta) \in X_T^\e$, ${\bf f} \in L^2(0,T;L^2(\W))$. Then, the functions $\Phi, \Upsilon, \Lambda$ defined as
\begin{eqnarray}
\label{phi}
 \Phi &=& \zeta - \beta \bw \cdot \nabla \psi - \psi^3-u \psi - \lambda \psi \bw^2
	\\
	\label{ups}
	\Upsilon &=& \mathcal{P}[{\bf f}-\k \div (\nabla\psi \otimes \nabla\psi) + \lambda \psi (\psi_t + \bw\cdot \nabla \psi) \bw 
	- (\nabla \bw) \bw]
	\\
	\label{lambda}
	\Lambda &=& -\psi_t - \bw \cdot \nabla \psi
\end{eqnarray}
satisfy
$\Phi \in L^2(0,T; H^1(\W)) \cap H^1(0, T; H^{1}(\W)')$, $\Upsilon \in L^2(0,T; L^2_{\rm div}(\W))$, $\Lambda \in L^2(0,T; L^2(\W))$.
\end{lem}

\noindent
{\it Proof.}
From definition (\ref{phi}) and inequalities (\ref{H1H2}) and (\ref{H2H2}), it follows that
\begin{eqnarray}\label{PhiH1}
\|\Phi\|_{H^1} &\leq&
C(\|\zeta\|_{H^1} + \|\bw \cdot \nabla \psi\|_{H^1} + \|\psi^3\|_{H^1} + \|\psi\|_{H^1} + \|\psi \bw^2\|_{H^1})
\nonumber\\
&\leq&
C(\|\zeta\|_{H^1}+ \|\bw\|_{H^2} \|\psi\|_{H^2} + \|\psi\|^2_{H^2}\|\psi\|_{H^1} + \|\psi\|_{H^1}
\nonumber\\
&&+ \|\psi\|_{H^2} \|\bw\|_{H^1}\|\bw\|_{H^2}).
\end{eqnarray}
The assumption $(\psi,\bw,\zeta) \in X_T^\e$ and Theorem \ref{thm-evans} guarantee that $\psi \in C(0, T;$ $\widehat H^2(\W))$, 
$\bw \in C(0,T; H^1_{\rm div}(\W))$, $\zeta \in C(0,T; \widehat H^1(\W))$. Accordingly, Young's inequality yields
$$
\int_0^T\|\Phi\|_{H^1}^2 \rmd t < \infty.
$$

In order to prove that $\Phi \in H^1(0,T;H^{1}(\W)')$, we differentiate equation (\ref{phi}) with respect to $t$ and we evaluate $\|\Phi_t\|_{(H^{1})'}$. In view of (\ref{-1norm}) and the regularity of the functions $\psi$, $\bw$, $\zeta$, we have
\begin{eqnarray}\label{fit-1}
&&\fl\|\Phi_t\|_{(H^{1})'} \leq
C(\|\zeta_t\|+ \|\bw_t\|\| \nabla \psi\|_{H^1} + \|\bw\|_{H^1}\| \nabla \psi_t\| + \|\psi\|_{H^1}^2\|\psi_t\| 
+ \|\psi_t\|\nonumber
\\
&&  + \|\psi_t\|\| \bw\|_{H^1}^2+ \|\psi\|_{H^1}\|\bw\|_{H^1} \|\bw_t\| ).
\end{eqnarray}
In addition, Young's inequality leads to
\bea
&&\fl\int_0^\tau \|\Phi_t\|^2_{(H^{1})'} \rmd t
\leq
C\int_0^\tau[\|\zeta_t\|^2 + (1+ \|\bw\|_{H^1}^4 +\|\psi\|_{H^1}^4)\|\psi_t\|^2_{H^1}  \\
&&+ (\|\psi\|_{H^1}^2 \|\bw\|_{H^1}^2+\|\psi\|_{H^2}^2) \|\bw_t\|^2 ]\rmd t<\infty.
\eea

Now we consider equation (\ref{ups}). We get
\bea
\fl\|\Upsilon\| \leq 
C(\|{\bf f}\|+\|\psi\|_{H^2}\|\psi\|_{H^3} + \|\psi\|_{H^1}\|\psi_t\|_{H^1}\|\bw\|_{H^1} + \|\psi\|_{H^2}^2\|\bw\|^2_{H^1}
+ \|\bw\|_{H^1}\|\bw\|_{H^2} ),
\eea
which implies $\Upsilon \in L^2(0,T ;L^2_{\rm div}(\W))$.

Similarly, from (\ref{lambda}) we deduce
$$
\|\Lambda\| \leq \|\psi_t\| + \|\bw\|_{H^1} \|\psi\|_{H^2},
$$
so that $\Lambda \in L^2(0,T ;L^2(\W))$.
\qed

\begin{theor}\label{exloc}
Suppose that $\vf_0 \in \widehat H^2(\W)$, $\bv_0 \in H^1_{\rm div}(\W)$, $\mu_0\in \widehat H^1(\W)$, ${\bf f} \in L^2(0,T;L^2(\W))$. Then, problem $(P_\e)$ admits a unique local solution for a sufficiently small time $\tau\in(0,T)$.
\end{theor}

\noindent{\it Proof.}
For any $\tau>0$, we define
\bea
\mathcal S : X_\tau ^\e &\to & X_\tau^\e
\\
(\psi, \bw, \zeta) &\mapsto& (\vf, \bv, \mu),
\eea
where $(\vf, \bv, \mu)$ is the unique solution of problem $(P_L)$ and $\Phi, \Upsilon, \Lambda$ are defined by (\ref{phi})-(\ref{lambda}). Thanks to Lemmas \ref{PL}-\ref{lemma-lambda}, $\mathcal S$  is well defined.
Furthermore, we consider the bounded subset $B_\tau\subset  X_\tau^\e$ that consists of the functions $(\psi, \bw, \zeta)$ satifying the following conditions:
\begin{eqnarray}
\label{m1}
	\int_0^\tau [\|\psi \|_{H^3}^2 + \|\bw \|_{H^2}^2 + \|\zeta \|_{H^2}^2 ]\rmd t\leq n_1, 
	\\
	\label{m2}
	\int_0^\tau [\|\psi_t \|_{H^1}^2 + \|\bw_t \|^2 + \|\zeta_t \|^2 ] \rmd t \leq n_2,
	\\
	\label{m3}
	\int_0^\tau \|\psi_{tt}\|^2_{(H^1)'} \rmd t \leq n_3,	
\end{eqnarray}
where $n_1,n_2,n_3$ are positive constants which will be specified in the sequel.
In particular, in view of Theorem \ref{thm-evans}, there exist suitable constants $C(n_1,n_2)$, $C(n_2,n_3)$ such that
\begin{eqnarray}\label{Mmax1}
&&\max_{[0,\tau]} [\|\psi\|_{H^2}^2 + \|\bw\|_{H^1}^2 + \|\zeta\|_{H^1}^2] \leq C(n_1,n_2), \\ 
\label{Mmax2}
&&\max_{[0,\tau]} \|\psi_t\|^2 \leq C(n_2,n_3).
\end{eqnarray}

The proof is divided into two steps.

\noindent
{\it 1. $\mathcal S$ maps $B_\tau$ in itself.}

\noindent
Throughout this proof we denote by $C$ a generic positive constant which is allowed to depend also on $\e$.
Let us consider equation (\ref{vf-lin}). 
By multiplying in $L^2(\W)$ by $\vf$ and integrating by parts, we obtain
$$
	\frac{\beta}{2}\frac{\rmd}{\rmd t} \|\vf\|^2  + \kappa\|\nabla\vf\|^2 
	\leq \eta \|\Phi\|^2 + C\|\vf\|^2,
$$
where $\eta$ is a (small) positive constant which will be chosen later.
From definition (\ref{phi}) of $\Phi$, the Sobolev embedding theorem (\ref{Sob1}) and relations (\ref{Mmax1})-(\ref{Mmax2}), we deduce the inequality
\bea
\fl\|\Phi\| \leq
C(\|\zeta\| + \|\bw\|_{H^1} \|\nabla \psi\|_{H^1} + \|\psi\|^3_{H^1} + \|\psi\| + \|\psi\|_{H^1} \|\bw\|^2_{H^1})
\leq C(n_1,n_2).
\eea
Hence, an application of Gronwall's inequality leads to the estimate
$$
\|\vf\|^2 \leq \rme^{C\tau} \left[ \|\vf_0\|^2 + \frac{2\eta}{\beta} \int_0^\tau \|\Phi\|^2\rmd t \right]\leq \rme^{C\tau} \left[ \|\vf_0\|^2 + C(n_1,n_2)\eta \tau \right].
$$
Choosing $n_1 > 2\|\vf_0\|^2$ and $\eta$ and $\tau$ small enough, we infer that $\|\vf\|^2 \leq n_1$.

Now we multiply (\ref{vf-lin}) in $L^2(\W)$ by $(\vf_t-\Delta\vf_t)$ and we integrate by parts. By taking Young's inequality into account, we obtain
\begin{equation}\label{stima:grad-vf}
	\frac{\kappa}{2}\frac{\rmd }{\rmd t} [\|\Delta\vf\|^2 + \|\nabla\vf\|^2] +\frac{ \beta}{2}\|\vf_t\|^2_{H^1}
	\leq C \|\Phi\|^2_{H^1}.
\end{equation}
From (\ref{PhiH1}), we deduce that
\begin{eqnarray}\label{phi-h1}
\fl\|\Phi\|_{H^1}^2 \leq C(\|\zeta\|^2_{H^1} + \|\bw\cdot\nabla\psi\|^2_{H^1} + \|\psi\|^2_{H^1}\|\psi\|^4_{H^2} + \|\psi\|^2_{H^1}+ \|\psi \bw^2\|^2_{H^1}). 
\end{eqnarray}
In particular,  in view of (\ref{Sob1}), (\ref{L4}), (\ref{infty}), (\ref{H1H2}) and Young's inequality, we have
\begin{eqnarray}\nonumber
\fl	\|\bw\cdot \nabla\psi\|^2_{H^1} 
	&\leq&
	C(\|\bw\cdot\nabla\psi\|^2 + \|(\nabla\bw)\nabla\psi\|^2 + \|(\nabla\nabla\psi)\bw\|^2)
	\\
	\nonumber
	&\leq&
	C(\|\bw\|^2_{H^1} \|\psi\|^2_{H^2} + \|\nabla\bw\|^2_4 \|\psi\|_{H^2}^2 + \|\nabla\nabla\psi\|_4^2 \|\bw\|^2_{H^1})
	\\
	\label{phi:1-2}
	&\leq&
	C (\|\bw\|^2_{H^1} \|\psi\|^2_{H^2} + \|\psi\|_{H^2}^4 \|\bw\|^2_{H^1} 
	+ \|\psi\|_{H^2}^2 \|\bw\|_{H^1}^4) 	\nonumber
	\\
	&&+ \eta \|\bw\|^2_{H^2}+ \eta\|\psi\|_{H^3}^2,
\end{eqnarray}
for any $\eta>0$.
By repeating similar arguments, we infer that
\begin{eqnarray}
\nonumber
\fl\|\psi \bw^2\|^2_{H^1} &\leq& C (\|\psi \bw^2\|^2 + \| \bw^2\nabla\psi\|^2 + \|\psi (\nabla\bw) \bw\|^2)
\\
\nonumber
& \leq &
C (\|\psi\|^2_{H^1} \|\bw\|^4_{H^1} + \|\nabla\psi\|_{H^1}^2 \|\bw\|^4_{H^1} 
 + \|\psi\|_{H^1}^2 \|\nabla\bw\|_4^2 \|\bw\|_{H^1}^2)\nonumber
\\
\label{phi:1-1}
&\leq &
C ( \|\psi\|_{H^2}^2 \|\bw\|^4_{H^1} + \|\psi\|_{H^1}^4 \|\bw\|_{H^1}^6) + \eta\|\bw\|_{H^2}^2.
\end{eqnarray}
Collecting (\ref{stima:grad-vf})-(\ref{phi:1-1}) and taking (\ref{m1}),(\ref{Mmax1}), into account, we prove the estimate
\bea
&&\fl\frac{\kappa}{2} [\|\Delta\vf\|^2 + \|\nabla\vf\|^2] + \frac{\beta}{2}\int_0^\tau\|\vf_t\|^2_{H^1} \rmd t\\
	&&\leq \frac{\kappa}{2} [\|\Delta\vf_0\|^2 + \|\nabla\vf_0\|^2]  
	+ 3\eta n_1 
+ C (n_1,n_2)\tau.
\eea
Thus, taking $n_2$ greater enough according to $\|\vf_0\|^2_{H^2}$ and $\tau$, $\eta$ small enough, we have
$$
\int_0^\tau \|\vf_t\|^2_{H^1}  \rmd t \leq n_2. 
$$
From (\ref{vf-lin}) it follows that
$$
\kappa\|\Delta\vf\|_{H^1} \leq \beta \|\vf_t\|_{H^1} + \|\Phi\|_{H^1}.
$$
Hence, on account of (\ref{H3}) and (\ref{phi-h1}) we deduce that
$$
\int_0^\tau \|\vf\|_{H^3}^2\leq n_1,
$$
with $\tau$ small enough and $n_1>n_2$.

Finally, we observe that by differentiating (\ref{vf-lin}) with respect to time, we obtain
$$
	\beta \vf_{tt} -\k \Delta \vf_t = \Phi_t,
$$
which implies
$$
\int_0^\tau\|\vf_{tt}\|^2_{(H^1)'} \rmd t\leq C\int_0^\tau[\|\vf_t\|^2_{H^1}+\|\Phi_t\|^2_{(H^1)'}]\rmd t.
$$
Owing to (\ref{fit-1})-(\ref{Mmax1}), we get
$$
\int_0^\tau\|\vf_{tt}\|^2_{(H^1)'} \rmd t\leq C (n_1,n_2) + C\tau.
$$
Choosing $\tau$ small enough and $n_3 > C (n_1, n_2)$, we deduce that
$$
\int_0^\tau\|\vf_{tt}\|^2_{(H^1)'} \rmd t\leq n_3.
$$

Now we multiply (\ref{v-lin}) by $(\bv  + A\bv)$ and we integrate over $\W$, thus obtaining
\begin{equation}\label{stima:grad-v}
	\frac{1}{2}\frac{\rmd }{\rmd t} \|\bv\|^2_{H^1} + \frac{\nu}{2} \|\nabla\bv\|^2 
	+ \frac{\nu}{2} \|A\bv\|^2 
	\leq C \|\Upsilon\|^2.
\end{equation}
The definition (\ref{ups}) of $\Upsilon$ implies the following inequality
\begin{eqnarray}
\fl\|\Upsilon\|^2 &\leq& C [\|{\bf f}\|^2 + \|\nabla\nabla\psi\|_4^2 \|\nabla\psi\|_4^2 + \|\psi\|_{H^1}^2 \|\psi_t\|_{4}^2 \|\bw\|_{H^1}^2\nonumber\\
&&+ \|\psi\|_{H^1}^2 \|\bw\|^4_{H^1} \|\psi\|^2_{H^2} + \|\nabla\bw\|_4^2 \|\bw\|_4^2]
\nonumber
\\
\label{ups_L2}
&\leq&
\eta \|\psi\|_{H^3}^2 + \eta \|\psi_t\|^2_{H_1} + \eta\|\bw\|_{H^2}^2 +C\|{\bf f}\|^2 + C(n_1,n_2,n_3),
\end{eqnarray}
for any $\eta>0$.
An integration of (\ref{stima:grad-v}) over $(0, \tau)$ leads to
\bea
&&\fl\|\bv\|^2_{H^1}+\nu \int_0^\tau  [\|\nabla\bv\|^2 	+ \|A\bv\|^2] \rmd t\\
	&&\leq \|\bv_0\|^2_{H^1}  + \eta( 2n_1 + n_2) + C(n_1, n_2, n_3)\tau+C\|{\bf f}\|^2_{L^2(0,\tau;L^2)} .
\eea
We take $n_1,n_2$ large enough and $\eta, \tau$ small enough. Accordingly,
\bea
\int_0^\tau \|\bv\|^2_{H^2} \rmd t &\leq & C\int_0^\tau (\|\bv\|^2 + \|A\bv\|^2 ) \rmd t \leq n_1,
\eea
and a comparison with (\ref{v-lin}) yields
$$
\int_0^\tau \|\bv_t\|^2 \rmd t \leq n_2.
$$

We multiply (\ref{mu-lin}) by $(\mu-\Delta \mu)$ and we integrate over $\W$, thus obtaining
\begin{equation}\label{stima:grad-mu}
	\frac{\e}{2}\frac{\rmd }{\rmd t} \|\mu \|^2_{H^1}+ \frac{\gamma}{2} \|\Delta\mu \|^2 
	+ \gamma \|\nabla \mu \|^2\leq C (\|\mu\|^2+\|\Lambda \|^2).
\end{equation}
In view of (\ref{lambda}), (\ref{Mmax1}) and (\ref{Mmax2}), we infer that
$$
\|\Lambda \|^2 \leq \|\psi_t\|^2 + C \|\bw\|^2_{H^1} \|\psi\|^2_{H^2}
\leq 
+ C(n_1,n_2, n_3).
$$
As a consequence, by applying Gronwall's inequality to (\ref{stima:grad-mu}) we obtain
$$
\frac{\e}{2} \|\mu \|_{H^1}^2 +\frac{\gamma}{2}\int_0^\tau  \|\Delta\mu \|^2 \rmd t \leq C \rme^{C\tau}\left[
	 \|\mu_0 \|^2_{H^1}+ C(n_1,n_2,n_3)\tau	\right].
$$
By choosing $n_1$ large enough and $\tau$ small enough, we deduce that
$$
\int_0^\tau \|\mu \|^2_{H^2} \rmd t \leq 
C\int_0^\tau (\|\mu \|^2 + \|\Delta\mu \|^2) \rmd t \leq n_1.
$$
A comparison with (\ref{mu-lin}) yields
$$
\e^2 \|\mu_t\|^2 \leq 2(\gamma^2 \|\Delta\mu\|^2 + \|\Lambda\|^2) 
\leq 
2\gamma^2 \|\Delta\mu\|^2 + C(n_1,n_2,n_3).
$$
Accordingly, we have
$$
\int_0^\tau \|\mu_t\|^2 \rmd t \leq n_2,
$$
provided that $n_2$ is large enough and $\tau$ is small enough.

\bigskip

\noindent
{\it 2. $\mathcal S$ is a contraction in $B_\tau$ if $\tau$ is small enough.}
Let $(\psi_1, \bw_1, \zeta_1), (\psi_2, \bw_2, \zeta_2) \in B_\tau$ and $(\vf_1, \bv_1, \mu_1)$, $(\vf_2, \bv_2, \mu_2)$ be the corresponding two solutions of the linear problem $(P_L)$ with the same initial data. We denote by $\psi=\psi_1-\psi_2$, $\bw=\bw_1 -\bw_2$, $\zeta= \zeta_1 - \zeta_2$ and prove that $\mathcal S:B_\tau\to B_\tau$ is a contraction mapping with respect to the metric induced by the norm
\bea
\fl|\|(\psi,\bw,\zeta)|\|^2&=&\|\psi\|_{L^2(0,\tau;H^3)}^2+\|\psi_t\|_{L^2(0,\tau;H^1)}^2+\|\psi_t\|^2_{C([0,\tau];L^2)}
+\|\bw\|_{L^2(0,\tau;H^2)}^2
\\
&&+\|\bw_t\|_{L^2(0,\tau;L^2)}^2+\|\zeta\|^2_{C([0,\tau];H^1)}.
\eea
It is worth noting that $X_\tau$ is a closed metric space with respect to the metric $|\| \cdot |\|$. 
Therefore, our aim consists in proving that
\begin{equation}\label{Scontr}
|\|\mathcal{S}(\psi_1, \bw_1, \zeta_1) - \mathcal{S}(\psi_2, \bw_2, \zeta_2) |\| \leq L_{\tau}
|\|(\psi, \bw, \zeta) |\|,
\end{equation}
with $0<L_{\tau}<1$.
From (\ref{vf-lin}) and (\ref{phi}), we deduce that $\vf = \vf_1 -\vf_2$ satisfies the following equation:
\begin{equation}\label{vf-contr}
	\beta \vf_t - \kappa \Delta\vf = 
	\Phi_1 - \Phi_2,
\end{equation}
where
\bea
	\fl\Phi_1 - \Phi_2 &=&
		\zeta - \beta (\bw_1\cdot \nabla\psi + \bw\cdot\nabla\psi_2) - \psi (\psi_1^2 + \psi_1 \psi_2 + \psi_2^2) - u\psi \\ &&- \lambda [\psi\bw_1^2 - \psi_2 (\bw_1 + \bw_2)] \cdot \bw.
\eea
Let us multiply (\ref{vf-contr}) by $\vf$ and integrate over $\W$. By means of (\ref{Sob1}) and Young's inequality, we obtain
\bea
&&\fl\frac{\b}{2}\frac{\rmd }{\rmd t}\|\vf\|^2+\k\|\nabla\vf\|^2
\leq C\|\vf\|^2+\eta\left[\|\zeta\|^2+(\|\psi_1\|^4_{H^1}+\|\psi_1\|^2_{H^1}\|\psi_2\|^2_{H^1}+\|\psi_2\|^4_{H^1}\right.
\\
&&+\|\bw_1\|^4_{H^1} +1 )\|\psi\|^2_{H^1}+(\|\psi_2\|_{H^2}^2+\|\psi_2\|^2_{H^1} \|\bw_1 + \bw_2\|^2_{H^1} 
 )\|\bw\|^2_{H^1}\left. 
 +\|\bw_1\|^2_{H^1} \|\psi\|_{H^2}^2
\right]
\eea
for every $\eta>0$.
The assumption  $(\psi_1, \bw_1, \zeta_1), (\psi_2, \bw_2, \zeta_2) \in B_\tau$ guarantees that
\bea
\frac{\b}{2}\frac{\rmd }{\rmd t}\|\vf\|^2+\k\|\nabla\vf\|^2
\leq C\|\vf\|^2+C\eta\left[\|\zeta\|^2+\|\psi\|^2_{H^2}+\|\bw\|^2_{H^1}\right].
\eea
Hence Gronwall's inequality leads to the estimate
$$
	\|\vf\|^2\leq C\eta \rme^{C\tau}\int_0^\tau \left[\|\zeta\|^2+\|\psi\|^2_{H^2}+\|\bw\|^2_{H^1}\right]\rmd t \leq
	C\eta \tau \rme^{C\tau}|\|(\psi,\bw,\zeta)|\|^2.
$$

We multiply (\ref{vf-contr}) in $L^2(\W)$ by $(\vf_t-\Delta\vf_t)$. An integration by parts and Young's inequality yield
\bea
\frac{\kappa}{2} \frac{\rmd }{\rmd t} (\|\Delta \vf \|^2  + \|\nabla \vf \|^2) +
\frac{\beta}{2} \|\vf_t \|^2_{H^1}
&\leq&
C \| \Phi_1 - \Phi_2 \|^2_{H^1}.
\eea
From the definition of $\Phi_1 - \Phi_2$ and inequalities (\ref{H1H2}), (\ref{H2H2}), we have
\bea
&&\fl\| \Phi_1 - \Phi_2 \|^2_{H^1} 
\leq
C\left[ \|\zeta \|^2_{H^1}+  \|\bw_1\cdot \nabla\psi\|^2_{H^1}+ \|\bw\cdot \nabla\psi_2\|^2_{H^1}\right.\\
 &&+(\|\psi_1\|_{H^2}^4 + \|\psi_1\|_{H^2}^2 \|\psi_2\|_{H^2}^2 + \|\psi_2\|^4_{H^2} +1) \|\psi\|_{H^1}^2 
\\
&&\left. 
 + \|\psi \bw_1^2\|^2_{H^1}+ \|\psi_2\|_{H^2}^2 \|(\bw_1+\bw_2)\cdot \bw\|^2_{H^1}
\right].
\eea
Owing to Sobolev embedding theorem and accounting for (\ref{Sob1}), (\ref{L4}) and (\ref{Mmax1}), we obtain
\bea
\fl\|(\bw_1+\bw_2)\cdot \bw\|^2_{H^1} \leq  C(\|\bw_1+\bw_2\|^2_{H^1} \|\bw\|^2_{H^1}\\
  + \|\nabla(\bw_1+\bw_2)\|^2_4 		\|\bw\|^2_{H^1}+ \|\bw_1+\bw_2\|^2_{H^1} \|\nabla\bw\|^2_{4})
\\
\leq 
C(\|\bw\|^2_{H^1} + \|\bw_1+\bw_2\|_{H^2} \|\bw\|^2_{H^1} + \|\bw\|_{H^1} \|\bw\|_{H^2})
\\
\leq  
C( 1 + \|\bw_1\|_{H^2} + \|\bw_2\|_{H^2} )\|\bw\|^2_{H^1} + \eta_1\|\bw\|^2_{H^2},
\eea
with $\eta_1>0$.
Moreover, proceeding as in the second inequalities of (\ref{phi:1-2}) and (\ref{phi:1-1}) we can prove the estimates
\bea
\|\psi \bw_1^2\|^2_{H^1}
&\leq&
C(\|\psi\|^2_{H^2}+\|\bw_1\|_{H^2}\|\psi\|^2_{H^1}),
\\
\|\bw_1\cdot \nabla\psi\|^2_{H^1} 
& \leq &
C(1+ \|\bw_1\|_{H^2}) \|\psi\|^2_{H^2} + \eta_2 \|\psi\|_{H^3}^2,
\\
\|\bw\cdot \nabla\psi_2\|^2_{H^1} 
& \leq &
C(1+ \|\psi_2\|_{H^3})\|\bw\|^2_{H^1}  + \eta_1\|\bw\|^2_{H^2},
\eea
where $\eta_1, \eta_2$ are suitable positive constants.
Collecting the previous results, we get
\bea
\fl\frac{\kappa}{2} \frac{\rmd }{\rmd t} [(\|\Delta \vf \|^2  + \|\nabla \vf \|^2) +
\frac{\beta}{2} \|\vf_t \|^2_{H^1}\leq  2\eta_1 \|\bw\|^2_{H^2} + \eta_2 \|\psi\|^2_{H^3}\\
+
C\left[\|\zeta \|^2_{C([0,\tau]; H^1)}  
+ (1 + \|\bw_1\|_{H^2}) \|\psi\|^2_{C([0,\tau]; H^2)}\right.\\
\left.
+ (1 +  \|\bw_1\| _{H^2} + \|\bw_2\|_{H^2} + \|\psi_2\|_{H^3} ) \|\bw\|^2_{C([0,\tau]; H^1)}
\right].
\eea
An integration over $(0,\tau)$, Theorem \ref{thm-evans} and H\"older's inequality yield
\bea
\nonumber
\fl
&&\frac{\kappa}{2} (\|\Delta \vf \|^2  + \|\nabla \vf \|^2) + \frac{\beta}{2} \int_0^\tau \|\vf_t \|^2_{H^1} \rmd t 
\leq 
2\eta_1 \|\bw\|^2_{L^2(0,\tau; H^2)} + \eta_2 \|\psi\|^2_{L^2(0,\tau; H^3)}\\
\nonumber
 &&+ 
C\tau |\|(\psi, \bw, \zeta) |\|^2+
C\sqrt{\tau}\, \|\bw_1\|_{L^2(0,\tau;H^2)} \|\psi\|^2_{C([0,\tau]; H^2)} 
\\
&&+ C\sqrt{\tau} \left(\|\bw_1\|_{L^2(0,\tau;H^2)}+ \|\bw_2\|_{L^2(0,\tau;H^2)} + \|\psi_2\|_{L^2(0,\tau;H^3)} \right) \|\bw\|^2_{C([0,\tau]; H^1)}.
\eea
Hence, with a suitable choice of $\eta_1, \eta_2$ and $\tau$, we have
\begin{eqnarray}\label{vf-nabla}
\frac{\kappa}{2} (\|\Delta \vf \|^2  + \|\nabla \vf \|^2) + \frac{\beta}{2} \int_0^\tau \|\vf_t \|^2_{H^1} \rmd t
\leq
L_{\tau} |\|(\psi,\bw,\zeta)|\|^2,
\end{eqnarray}
where $0<L_{\tau}<1$.
A comparison with (\ref{vf-contr}) and Young's inequality yield
$$
\|\Delta\vf\|_{H^1}^2 \leq C(\|\vf_t\|_{H^1}^2 + \|\Phi_1 - \Phi_2\|_{H^1}^2),
$$
which, in view of (\ref{H3}) and (\ref{vf-nabla}), guarantees
\bea
\int_0^\tau\|\vf\|_{H^3}^2\rmd t
\leq 
C\int_0^\tau (\|\vf\|^2_{H^1} +\|\Delta\vf\|_{H^1}^2) \rmd t 
\leq 
L_{\tau} |\|(\psi, \bw, \zeta) |\|^2.
\eea

From (\ref{v-lin}) and (\ref{ups}) it follows that
\begin{equation}\label{v-contr}
	\bv_t + \nu A\bv  = \Upsilon_1 - \Upsilon_2,
\end{equation}
where
\bea
\fl\Upsilon_1 - \Upsilon_2 &=& \mathcal{P}[-\kappa \div(\nabla\psi_1 \otimes \nabla\psi + \nabla\psi \otimes \nabla\psi_2)
+\lambda \psi_1\psi_{1t}\bw + \lambda \psi_1\psi_{t}\bw_2+ \lambda \psi\psi_{2t}\bw_2
\\
&&+\lambda \psi_1(\bw_1\cdot\nabla\psi_{1})\bw
+\lambda \psi(\bw_1\cdot\nabla\psi_{1})\bw_2 + \lambda \psi_2(\bw\cdot\nabla\psi_{1})\bw_2 \\
&&+\lambda \psi_2(\bw_2\cdot\nabla\psi)\bw_2 -(\nabla\bw)\bw_1 - (\nabla\bw_2)\bw].
\eea

Let us multiply (\ref{v-contr}) by $(\bv +A\bv)$ thus obtaining
\bea
\frac{1}{2}\frac{\rmd }{\rmd t}[\|\bv\|^2 + \nu\|\nabla\bv\|^2] + \frac{\nu}{2}(\|A\bv\|^2 + \|\nabla\bv\|^2)
\leq
C\|\Upsilon_1 - \Upsilon_2\|^2.
\eea
By means of the Sobolev embedding theorems and (\ref{m1}), (\ref{m2}) inequalities, the same arguments used to prove (\ref{ups_L2}) lead to the inequality
\bea
\fl\|\Upsilon_1 - \Upsilon_2\|^2 
&\leq&
 C[(\|\psi_1\|_{H^3} + \|\psi_2\|_{H^3} + 1)\|\psi\|^2_{H^2}
 + \|\psi_t\|^2+ \|\psi_{2t}\|_{H^1} \|\psi\|^2_{H^1}
\\
&&+ (\|\psi_{1t}\|_{H^1}+ \|\bw_2\|_{H^2}+1) \|\bw\|^2_{H^1}]
+ \eta_1 \|\bw\|_{H^2}^2 
 + \eta_2\|\psi\|_{H^3}^2 
+ \eta_3\|\psi_t\|_{H^1}^2,
\eea
for any $\eta_1,\eta_2, \eta_3>0$.
An integration over $(0,\tau)$ leads to
\begin{eqnarray*}
\nonumber
&&\fl
\frac{1}{2}[\|\bv\|^2 + \nu\|\nabla\bv\|^2]  + \frac{\nu}{2}\int_0^\tau(\|A\bv\|^2 + \|\nabla\bv\|^2)\rmd t
\\
\nonumber
&&\leq
C\sqrt{\tau}\left[\|\psi_1\|_{L^2(0,\tau;H^3)} + \|\psi_2\|_{L^2(0,\tau;H^3)}\right.\\
&&\left. + \|\psi_{2t}\|_{L^2(0,\tau;H^1)}\right] \|\psi\|^2_{C(0,\tau;H^2)}
\\
\nonumber
 &&+
C\sqrt{\tau}\left[\|\psi_{1t}\|_{L^2(0,\tau;H^1)}+ \|\bw_2\|_{L^2(0,\tau;H^2)}\right]\|\bw\|^2_{C(0,\tau;H^1)}
\\
 &&+
C\tau \left[\|\psi\|^2_{C(0,\tau;H^2)} +\|\psi_t\|^2_{C(0,\tau;L^2)}+ \|\bw\|^2_{C(0,\tau;H^1)} \right]\\
&&+ \eta_1 \|\bw\|_{L^2(0,\tau;H^2)}^2 + \eta_2 \|\psi\|_{L^2(0,\tau;H^3)}^2
+\eta_3 \|\psi_t\|^2_{L^2(0,\tau;H^1)}.
\end{eqnarray*}
Hence
\begin{equation}\label{v-H2}
\frac{1}{2}[\|\bv\|^2 + \nu\|\nabla\bv\|^2]  + \frac{\nu}{2}\int_0^\tau(\|A\bv\|^2 + \|\nabla\bv\|^2)\rmd t\leq L_\tau|\|(\psi,\bw,\zeta)\||.
\end{equation}
From (\ref{mu-lin}) and (\ref{lambda}), we obtain
\begin{equation}\label{mu-contr}
	\e \mu_t - \gamma \Delta \mu = \Lambda_1 - \Lambda_2,
\end{equation}
with
$$
\Lambda_1 - \Lambda_2 =-\psi_t - \bw_1 \cdot \nabla\psi - \bw \cdot \nabla\psi_2.
$$
We multiply (\ref{mu-contr}) by $(\mu -\Delta\mu)$ and we integrate over $\W$:
\bea
\fl\frac{\e}{2}\frac{\rmd }{\rmd t} \|\mu\|^2_{H^1} + \gamma\|\Delta\mu\|^2 + \gamma \|\nabla\mu\|^2\\
\leq C\|\mu\|^2_{H^1}+
\eta(\|\psi_t\|^2_{H^1} + \|\bw_1\|_{H^2}^2\|\psi\|_{H^2}^2 + \|\psi_2\|_{H^3}^2\|\bw\|^2_{H^1}).
\eea
Gronwall's inequality yields
\begin{eqnarray}
&&\fl\frac{\e}{2} \|\mu\|^2_{H^1} +\gamma \int_0^\tau \|\Delta\mu\|^2 \rmd t
	\leq
	\eta \rme^{C\tau} \left[\|\psi_t\|^2_{L^2(0,\tau;H^1)}+ \|\bw_1\|_{L^2(0,\tau;H^2)}^2\|\psi\|_{C([0,\tau];H^2)}^2\right.
	\nonumber\\
&&	\left.  + \|\psi_2\|^2_{L^2(0,\tau;H^3)}\|\bw\|^2_{C([0,\tau];H^1)}\right]
	\leq L_\tau |\|(\psi,\bw,\zeta)|\|^2.
\label{mu-H2}
\end{eqnarray}

Therefore, collecting (\ref{vf-nabla}), (\ref{v-H2}), (\ref{mu-H2}) and choosing $\tau$, $\eta_1,\eta_2, \eta_3$ small enough, we prove that
$$
\|\varphi\|_{L^2(0,\tau;H^3)}^2+\|\vf_t\|_{L^2(0,\tau;H^1)}^2+\|\bv\|_{L^2(0,\tau;H^2)}^2+\|\mu\|^2_{C([0,\tau];H^1)}\leq L_\tau |\|(\psi,\bw,\zeta)|\|^2.
$$
The control on the  remaining norms  $\|\vf_t\|_{C([0,\tau];L^2)}$ and $\|\bv\|_{L^2(0,\tau;L^2)}$  is obtained by comparison with (\ref{vf-contr}) and (\ref{v-contr}) respectively. Therefore (\ref{Scontr}) is proved.

\medskip

By means of a fixed point argument, the previous steps allow to prove that problem $(P_\e)$ admits a unique local solution $(\vf,\bv,\mu)$ in $X_\tau$, provided that $\tau$ is small enough.
\qed

\begin{lem} \label{lemma:apriori}
Any solution $(\vf^\e,\bv^\e,\mu^\e)$ of problem $(P_\e)$ satisfies the estimates 
\begin{eqnarray}\label{aprioriI}
&&	\fl\|\bv^\e(t)\|^2 + \|\vf^\e(t)\|^2_{H^1}+ \e\|\mu^\e(t)\|^2\nonumber\\
	&&+ \int_0^t \left[ \|\nabla\mu^\e\|^2 + \|\bv^\e\|^2_{H^1} +\|\vf_t^\e\|^2 +\|\vf^\e\|_{H^2}^2\right] \rmd x \leq C_0(T),
\\
\label{aprioriIII}
&&\fl\e\|\mu^\e(t)\|_{H^1}^2+\int_0^t[\|\mu^\e\|^2_{H^2}+\e\|\mu_t\|^2]\rmd t\leq C_0(T),
\\
\label{aprioriII}
&&\fl\|\Delta\vf^\e(t)\|^2+\|\nabla\bv^\e(t)\|^2\nonumber\\
&&+\int_0^t[\|\nabla\vf_t^\e\|^2+\|\vf\|_{H^3}^2 +\|A\bv^\e\|^2+\|\bv_t^\e\|^2]\rmd t
\leq C_0(T).
\end{eqnarray}
for any $t\in (0,T)$, where $C_0(T)$ is a positive constant depending on the inital data, the source ${\bf f}$ and the time $T$.
\end{lem}
{\bf Proof.}
By multiplying (\ref{weak1}) by $\vf_t^\e+\bv^\e\cdot\nabla\vf^\e$, we deduce the equality
\begin{eqnarray} \label{ap3}
\fl\frac{\k}{2}\frac{\rmd }{\rmd t}\|\nabla\vf^\e\|^2+\beta \|\vf_t^\e+\bv^\e\cdot\nabla\vf^\e\|^2
-\k\int_{\Omega}  (\bv^\e\cdot\nabla\vf^\e)\Delta\vf^\e \rmd x\nonumber\\
+\int_{\Omega} [(\vf^{\e})^3+u\vf^\e + \l (\bv^\e)^2 \vf^\e - \mu^\e](\vf^\e_t+\bv^\e\cdot\nabla\vf^\e) \rmd x =0.
\end{eqnarray}
Now let us multiply (\ref{weak2}) by $\bv^\e$
\begin{eqnarray}\label{ap2}
\fl\frac12 \frac{\rmd }{\rmd t}\|\bv^\e\|^2 +\nu\|\nabla\bv^\e\|^2+\int_{\Omega} \left[   \k\div(\nabla\vf^\e \otimes \nabla\vf^\e)\cdot\bv^\e\right. \nonumber\\
\left.-\l\vf^\e( \vf^\e_t+\bv^\e\cdot\nabla\vf^\e)(\bv^\e)^2 - {\bf f}\cdot \bv^\e \right] \rmd x =0.
\end{eqnarray}

Finally, we multiply (\ref{weak3}) by $\mu^\e$ and integrate by parts, thus obtaining
\begin{eqnarray}\label{ap1}
\frac{\e}{2}\frac{\rmd }{\rmd t}\|\mu^\e\|^2+\g\|\nabla\mu^\e\|^2+\int_{\Omega} (\vf^\e_t+\bv^\e\cdot\nabla\vf^\e)\mu^\e \rmd x =0.
\end{eqnarray}

It is easy to show that since $\bv^\e\in H^1_{\rm div}(\W)$, the following identity holds
$$
\int_\W\div(\nabla\vf^\e \otimes \nabla\vf^\e)\cdot\bv^\e \rmd x=\int_\W (\bv^\e\cdot\nabla\vf^\e)\Delta\vf^\e \rmd x.
$$
Therefore summing up equations (\ref{ap3})-(\ref{ap1}), we have
\begin{eqnarray}\label{ap3b}
&&\fl\frac12 \frac{\rmd }{\rmd t}\left[\k\|\nabla\vf^\e\|^2+\|\bv^\e\|^2+\e\|\mu^\e\|^2+\frac{1}{2}\|(\vf^\e)^2+u\|^2\right]+\beta \|\vf^\e_t+\bv^\e\cdot\nabla\vf^\e\|^2+\nu\|\nabla\bv^\e\|^2\nonumber\\
 &&+\g\|\nabla\mu^\e\|^2
=-\int_{\Omega} [(\vf^\e)^3+u\vf^\e]\bv^\e\cdot\nabla\vf^\e \rmd x + \int_{\W}{\bf f}\cdot \bv^\e \rmd x.
\end{eqnarray}
The first integral in the right hand side vanishes as a consequence of the identity
$$
\int_{\Omega} [(\vf^\e)^3+u\vf^\e]\bv^\e\cdot\nabla\vf^\e \rmd x=\int_{\Omega} \nabla\left[\frac{1}{4}(\vf^\e)^4+\frac{u}{2}(\vf^\e)^2\right]\cdot\bv^\e \rmd x
$$
and by applying the divergence theorem. 
Therefore, an integration of (\ref{ap3b}) over $(0,t)$ and Poincar\'e inequality provide
\begin{eqnarray}\label{ap4}
\fl\frac{1}{2}\left[\k\|\nabla\vf^\e(t)\|^2+\|\bv^\e(t)\|^2+\e\|\mu^\e(t)\|^2 + \frac12\|\vf^\e(t)^2+u\|^2\right]\nonumber
\\
 + \int_0^t \left[\beta\|\vf^\e_t+\bv^\e\cdot\nabla\vf^\e\|^2 + \frac{\nu}{2}\|\nabla\bv^\e\|^2 + \g \|\nabla\mu^\e\|^2 \right] \rmd t \leq C_0,
\end{eqnarray}
where $C_0>0$ depends on the norms of the initial data $\|\mu_0\|, \|\bv_0\|, \|\vf_0\|_{H^1}$ and of the source $\|{\bf f}\|$. In addition, an application of Young's inequality leads to
\begin{equation}\label{ap5}
\|\vf^\e(t)\|^2 \leq C(\|\vf^\e(t)^2+u\|^2 + 1) \leq C_0,
\end{equation}
for all $t\in[0,T]$.

Now we multiply (\ref{weak1}) by $\Delta\vf^\e$. An integration by parts, H\"older's and Young's inequalities imply
\begin{eqnarray}
&&\fl\int_0^t \|\Delta\vf^\e\|^2 \rmd t \leq C \int_0^t [\|\vf_t^\e+\bv^\e\cdot\nabla\vf^\e\|^2 +\|\nabla\mu^\e\|^2 + \|\nabla\vf^\e\|^2+ \|(\vf^\e)^3+u\vf^\e\|^2\nonumber\\
 &&+ \|\vf^\e(\bv^\e)^2\|^2  ]\rmd t.
\end{eqnarray}
In view of Sobolev embedding theorem we obtain
\begin{equation}\label{uvf}
	\|(\vf^\e)^3+u\vf^\e\|^2 + \|\vf^\e(\bv^\e)^2\|^2\leq C(\|\vf^\e\|_{H^1}^6+\|\vf^\e\|^2+\|\vf^\e\|_{H^1}^2\|\bv^\e\|^4_{6})
\end{equation}
and by means of (\ref{L6}), (\ref{ap4}) and (\ref{ap5}), last term of (\ref{uvf}) can be estimated as
$$
\|\vf^\e\|_{H^1}^2\|\bv^\e\|^4_{6} \leq C_0 \|\bv^\e\|^{8/3} \|\bv^\e\|_{H^1}^{4/3} \leq C_0 (\|\bv^\e\|^{8}+ \|\bv^\e\|_{H^1}^{2})
\leq C_0 (1+  \|\nabla\bv^\e\|^{2}).
$$
Therefore, on account of (\ref{ap4})-(\ref{uvf}), we have
$$
\int_0^t \|\vf^\e\|^2_{H^2} \rmd t \leq C \int_0^t (\|\vf^\e\|^2 + \|\Delta\vf^\e\|^2) \rmd t \leq C_0(T).
$$
In addition the following inequality holds
\begin{eqnarray*}
\fl\int_0^t \|\vf_t^\e\|^2 \rmd t
&\leq &
C \int_0^t (\|\vf_t^\e +\bv^\e \cdot \nabla\vf^\e\|^2 + \|\bv^\e \cdot \nabla\vf^\e\|^2) \rmd t\\
&\leq&
C_0\int_0^t ( 1 + \|\bv^\e\|^2_{4} \|\nabla\vf^\e\|^2_{4} )\rmd t\\
&\leq& C_0\int_0^t ( 1 + \|\bv^\e\|^2_{H^1}+ \|\nabla\vf^\e\|^2_{H^1} )\rmd t\leq C_0.
\end{eqnarray*}
Thus (\ref{aprioriI}) is proved.

By multiplying (\ref{weak3}) in $L^2(\W)$ by $-\Delta\mu^\e$, we obtain
$$
\frac{\e}{2}\frac{\rmd }{\rmd t}\|\nabla\mu^\e\|^2+\gamma\|\Delta\mu^\e\|^2\leq \int_\W (\vf_t^\e+\bv^\e\cdot\nabla\vf^\e) \Delta\mu^\e \rmd x\leq C\|\vf_t^\e+\bv^\e\cdot\nabla\vf^\e\|^2+\frac{\gamma}{2}\|\Delta\mu^\e\|^2.
$$
Hence,
\begin{equation}\label{mueps}
	\fl\e\|\nabla\mu^\e(t)\|^2+\gamma\int_0^t \|\Delta\mu^\e\|^2\rmd t \leq 
	\e\|\nabla\mu_0\|^2+ C\int_0^t \|\vf_t^\e+\bv^\e\cdot\nabla\vf^\e\|^2\rmd t  \leq C_0(T),
\end{equation}
where last inequality follows from (\ref{ap4}). 

A comparison with (\ref{weak3}) provides
\begin{equation}\label{mua}
\e\|\mu_t^\e\|^2\leq C(\|\Delta\mu^\e\|^2+\|\vf^\e_t+\bv^\e\cdot\nabla\vf^\e\|^2).
\end{equation}
Finally we observe that (\ref{weak1}) implies
\begin{equation}
\label{mub}
\|\mu^\e\|^2\leq C_0(\|\vf_t^\e\|^2+\|\vf^\e\|_{H^2}^2+\|\bv^\e\|^2_{H^1}).
\end{equation}
Collecting (\ref{mueps})-(\ref{mub}), we get (\ref{aprioriIII}).

In order to prove (\ref{aprioriII}) let us multiply (\ref{weak1}) by $-\Delta\vf_t$, (\ref{weak2}) by $A\bv$ and integrate over $\W$. By means of Young's inequality we get
\begin{eqnarray}\label{lapl}
&&\fl\frac{1}{2}\frac{\rmd }{\rmd t }[\k\|\Delta\vf^\e\|^2+\|\nabla\bv^\e\|^2]+\frac{\beta}{2}\|\nabla\vf^\e_t\|^2+\frac{\nu}{2}\|A\bv^\e\|^2
\nonumber
\\
&&\leq C\left[\|\nabla\mu^\e\|^2+\|(\nabla\bv^\e)\nabla\vf^\e\|^2+\|(\nabla\nabla\vf^\e)\bv^\e\|^2+\|(\vf^\e)^2\nabla\vf^\e\|^2\right.\nonumber\\
&&+\|\nabla\vf^\e\|^2+\|\nabla\vf^\e(\bv^\e)^2\|^2+\|\vf^\e(\nabla\bv^\e)\bv^\e\|^2+\|(\nabla\nabla\vf^\e)\nabla\vf^\e\|^2\nonumber\\
&&\left.+\|\vf^\e\vf^\e_t\bv^\e\|^2+\|\vf^\e(\bv^\e\cdot\nabla\vf^\e)\bv^\e\|^2 + \|{\bf f}\|^2 \right].
\end{eqnarray}
Owing to (\ref{L6}) and (\ref{aprioriI}), some of the terms of the right-hand side can be estimated as
\begin{eqnarray}\label{unif1}
\fl\|(\vf^\e)^2\nabla\vf^\e\|^2 +\|\nabla\vf^\e\|^2 + \|\nabla\vf^\e(\bv^\e)^2\|^2
\leq 
C_0(T)(\|\vf^\e\|_{H^2}^2+\|\bv^\e\|_{H^1}^2).
\end{eqnarray}
Similarly the remaining terms can be controlled as
\begin{eqnarray}\label{unif2}
&&\fl \|\vf^\e(\bv^\e\cdot\nabla\vf^\e)\bv^\e\|^2\leq C\|\bv^\e\|_6^4\|\vf^\e\|_{H^1}^2\|\nabla\vf\|_{H^1}^2
 \leq \chi(t)(\|\bv^\e\|^2_{H^1}+\|\vf^\e\|^2_{H^2})
 \\
&& \fl\|\vf^\e(\nabla\bv^\e)\bv^\e\|^2+ \|(\nabla\bv^\e)\nabla\vf^\e\|^2
 \leq C \|\nabla\bv^\e\|_4^2(\|\vf^\e\|_{H^1}^2\|\bv^\e\|_6^2+\|\nabla\vf^\e\|_4^2)
\nonumber\\
&&\leq \eta \|\bv^\e\|_{H^2}^2+\chi(t)( \|\bv^\e\|_{H^1}^2+1)
\\
&&\fl
\|(\nabla\nabla\vf^\e)\bv^\e\|^2+ \|(\nabla\nabla\vf^\e)\nabla\vf^\e\|^2\leq 
C \|\nabla\nabla\vf^\e\|_4^2(\|\bv^\e\|_4^2+\|\nabla\vf^\e\|^2_4)
\nonumber\\
\label{unif3} 
&&\leq
 \eta\|\vf^\e\|_{H^3}^2+\chi(t)(\|\bv^\e\|^2_{H^1}+\|\vf^\e\|_{H^2}^2+1)
 \\ \label{unif4}
&&\fl\|\vf^\e\vf^\e_t\bv^\e\|^2\leq C\|\vf^\e\|_{H^1}^2\|\vf^\e_t\|_4^2\|\bv^\e\|_6^2
\leq
\eta\|\nabla\vf^\e_t\|^2+\chi(t)(\|\bv^\e\|^2_{H^1}+1),
\end{eqnarray}
where $\chi$ is a $L^1$ function of time and $\eta$ is a suitable positive constant.

In order to evaluate the $H^3-$norm of $\vf^\e$, let us take the gradient of (\ref{weak1}) and obtain
\bea
&&\fl\|\nabla\Delta\vf^\e\|\leq C\left[\|\nabla\vf^\e_t\|+\|\nabla\mu^\e\|+\|\nabla\bv^\e\|_4\|\nabla\vf^\e\|_4+\|\bv^\e\|_4\|\nabla\nabla\vf^\e\|_4
\right.\\
&&+\|\vf^\e\|_6^2\|\nabla\vf^\e\|_6+\|\nabla\vf^\e\|
+\|\nabla\vf^\e\|_6\|\bv^\e\|_6^2
\left.+\|\vf^\e\|_{H^1}\|\nabla\bv^\e\|_4\|\bv^\e\|_6\right].
\eea
Interpolation inequalities (\ref{L4}), (\ref{L6}) and estimates (\ref{aprioriI})-(\ref{aprioriIII}) imply
\bea
&&\fl\|\nabla\Delta\vf^\e\|\leq\eta_1\|\bv^\e\|_{H^2}+\eta_2\|\vf^\e\|_{H^3}+
C[\|\nabla\vf^\e_t\|+ \|\nabla\mu^\e\| ]\\
&&+
C_0\left[1+\|\bv^\e\|_{H^1}\|\vf^\e\|_{H^2} 
+\|\vf^\e\|_{H^2}
+\|\bv^\e\|_{H^1}+\|\bv^\e\|_{H^1}^{5/3}\right].
\eea
with $\eta_1,\eta_2>0$.
Choosing suitably $\eta_2$ and owing to (\ref{H3}), (\ref{aprioriI}), we prove
\begin{equation}\label{phiH3}
\|\vf^\e\|_{H^3}^2\leq C [\|\nabla\vf^\e_t\|^2+\|\nabla\mu^\e\|^2 ]+\chi(t)[1+\|\bv^\e\|_{H^1}^2]+\eta_1\|\bv^\e\|_{H^2}^2,
\end{equation}

Substitution of (\ref{unif1})-(\ref{unif4}) into (\ref{lapl}), use of (\ref{aprioriI}), (\ref{mueps}) and (\ref{phiH3}) provide
\bea
&&\fl\frac{1}{2}\frac{\rmd }{\rmd t }[\k\|\Delta\vf^\e\|^2+\|\nabla\bv^\e\|^2] + \frac{\beta}{4}\|\nabla\vf_t^\e\|^2+\frac{\nu}{4}\|A\bv^\e\|^2+C\|\vf\|^2_{H^3}\\
\leq 
&&\chi(t)(\|\Delta\vf^\e\|^2+\|\nabla\bv^\e\|^2+1).
\eea
Gronwall's inequality and comparison with (\ref{weak2}) yield (\ref{aprioriII}).\qed

\begin{prop}
Problem $(P_\e)$ admits at least a solution $(\vf^\e,\bv^\e,\mu^\e)\in X_T$.
\end{prop}
{\it Proof.}
Theorem \ref{exloc} guarantees existence of a solution $(\vf^\e,\bv^\e,\mu^\e)$ defined in a small time interval $(0,\tau)$.
In order to extend this solution to the whole interval $(0,T)$ we need the following uniform estimate of the solution
\begin{equation}\label{uniform}
\|\vf^\e(t)\|_{H^2}+\|\bv^\e(t)\|_{H^1}+ \|\mu^\e(t)\|_{H^1}\leq K, \qquad\qquad  t\in[0,T],
\end{equation}
where $K$ is a positive constant depending only on the global data $\vf_0,\bv_0,\mu_0$ and $\e$, but independent of $t$.
Inequality (\ref{uniform}) is ensured by the estimates (\ref{aprioriI})-(\ref{aprioriII}) of Lemma \ref{lemma:apriori}.
Therefore, by applying Theorem \ref{exloc}, after a finite number of steps we find a global solution of problem $(P_\e)$ in $X_T$.\qed
\section{Well posedness of the original system}\label{sec:original}
\subsection{Existence of solutions}

\begin{theor}
Let $\vf_0 \in \hat H^2(\W)$, $\bv_0 \in H^1_{\rm div}(\W)$, $\mu_0 \in \hat H^1(\W)$. Then for any $T>0$, there exists at least a solution $(\vf,\bv,\mu) \in X_T$ of prolem $(P)$.
\end{theor}

\noindent
{\bf Proof.}
Let $(\vf^\e,\mu^\e,\bv^\e)\in X_T^\e$ be a global solution of problem $(P^\e)$.
From the a priori estimates of Lemma \ref{lemma:apriori}, we deduce that
\bea
\vf^\e &&\mbox{ is uniformly bounded in }  L^2(0,T,H^3(\W))\cap H^1(0,T,H^1(\W))\\
\bv^\e &&\mbox{ is uniformly bounded in }  L^2(0,T,H^2(\W))\cap H^1(0,T,L^2(\W))\\
\mu^\e &&\mbox{ is uniformly bounded in } L^2(0,T,H^2(\W))\\
\sqrt{\e}\mu^\e &&\mbox{ is uniformly bounded in } H^1(0,T,L^2(\W)).
\eea
As a consequence there exists a subsequence, denoted also $(\vf^\e,\bv^\e,\mu^\e)$ such that
\bea
\vf^\e \to \vf &&\mbox{\ weakly in }  L^2(0,T,H^3(\W))\\
\vf_t^\e \to \vf_t &&\mbox{\ weakly in } L^2(0,T,H^1(\W))\\
\bv^\e \to \bv &&\mbox{\ weakly in } L^2(0,T,H^2(\W))
\\
\bv^\e_t \to \bv_t &&\mbox{\ weakly in } L^2(0,T,L^2(\W))\\
\mu^\e \to \mu &&\mbox{\ weakly in } L^2(0,T,H^2(\W))\\
\e\mu^\e_t \to 0 &&\mbox{\ strongly in } L^2(0,T,L^2(\W)).
\eea
as $\e\to 0$.
In particular, as a consequence of  Aubin's theorem, we have
\bea
\vf^\e \to \vf &&\mbox{\ strongly in }  L^2(0,T,H^2(\W))
\\
\bv^\e \to \bv &&\mbox{\ strongly in } L^2(0,T,H^1(\W)).
\eea
The previous convergences allow us to pass to the limit as $\e\to 0$ into (\ref{weak1})-(\ref{weak3}) and to obtain (\ref{wk1})-(\ref{wk3}).
\qed

\subsection{Uniqueness of solution}
\begin{theor}
Problem $(P)$ admits a unique solution $(\vf,\bv,\mu) \in X_T$.
\end{theor}

\noindent
{\it Proof.}
First we notice that by repeating the same arguments of Lemma \ref{lemma:apriori} with $\e=0$, one can easily show that any solution $(\vf,\bv,\mu)$ of problem $(P)$ satisfies the following estimate
\begin{equation}\label{apr}
	\|\bv(t)\|^2_{H^1} + \|\vf(t)\|^2_{H^2}+  \int_0^t \left[  \|\bv\|^2_{H^2} +\|\vf_t\|^2_{H^1} \right] \rmd t  \leq C_0(T).
\end{equation}

Let $(\vf_1,\bv_1,\mu_1), (\vf_2,\bv_2,\mu_2) \in X_T$ be two solutions to problem $(P)$ with the same initial data and source ${\bf f}$.
We consider the differences
$$
\vf = \vf_1 - \vf_2, \qquad \bv= \bv_1 -\bv_2, \qquad \mu=\mu_1 -\mu_2,
$$
satisfying the following problem
\begin{eqnarray}
\label{diff-phi}
&&\fl\varphi_t = - \bv \cdot \nabla\vf_1 - \bv_2 \cdot \nabla\vf + \gamma \Delta\mu
\\
\label{diff-v}
&&\fl\bv_t =
- \nu A\bv - \mathcal{P} [ 
\kappa \div (\nabla\vf_1 \otimes \nabla\vf + \nabla\vf \otimes \nabla\vf_2)+\lambda \vf_1 (\vf_{1t} + \bv_1\cdot \nabla \vf_1) \bv_1
\nonumber\\
&&  - 
\lambda \vf_2 (\vf_{2t} + \bv_2\cdot \nabla \vf_2) \bv_2
+(\nabla \bv) \bv_1 + (\nabla \bv_2) \bv ]
\end{eqnarray}
with
\begin{eqnarray}\label{diff_mu}
&&\fl	\mu =\beta \vf_t  -\kappa \Delta\vf+  \beta \bv_1 \cdot \nabla \vf + \beta \bv \cdot \nabla \vf_2 + (\vf^2_1 + \vf^2_2 + \vf_1\vf_2)\vf + u\vf+\lambda \vf \bv_1^2 \nonumber\\
	 &&+ \lambda\vf_2 (\bv_1 + \bv_2) \cdot \bv. 
	\end{eqnarray}
We append to (\ref{diff-phi})-(\ref{diff_mu}) the initial conditions
$$
\vf(x,0)=0, \qquad \bv(x,0)=\bzero, \qquad a.e.\ x\in \W.
$$

We multiply equation (\ref{diff-phi}) by $\vf$ and we integrate over $\W$. Thus, we obtain
\bea
\frac12 \frac{\rmd }{\rmd t } \|\vf\|^2 = - \int_{\W} \left[(\bv \cdot \nabla\vf_1 + \bv_2 \cdot\nabla\vf)\vf + \gamma\nabla\mu \cdot \nabla \vf  \right]\rmd x.
\eea
By means of the H\"older and Young inequalities and in view of (\ref{apr}), we have
\begin{equation}\label{diff_Sphi}
	\frac12 \frac{\rmd }{\rmd t } \|\vf\|^2 \leq \|\bv\|^2 +  \frac{\gamma}{8} \|\nabla\mu\|^2 + C \|\vf\|^2_{H^1}.
\end{equation}
Now let us multiply (\ref{diff-phi})  in $L^2(\W)$ by $\mu$, thus obtaining
\begin{equation}\label{molt_mu}
	\int_\W\left[\vf_t\mu+(\bv \cdot \nabla\vf_1 + \bv_2 \cdot\nabla\vf)\mu + \gamma|\nabla\mu|^2\right]\rmd x=0.
\end{equation}
The first term of the integral may be rewritten by substituting expression (\ref{diff_mu}) as
\bea
&&\fl\int_\W \vf_t \mu \rmd x = \frac{\kappa}{2}\frac{\rmd }{\rmd t }\|\nabla\vf\|^2 + \b \|\vf_t\|^2 
+ \int_\W [\beta (\bv \cdot \nabla \vf_1 + \bv_2\cdot \nabla\vf) +\vf(\vf_1^2 +\vf_2^2 + \vf_1\vf_2)
\\
&&+ u\vf+ \l \vf\bv_1^2 + \l \vf_2(\bv_1+\bv_2)\cdot \bv ]\vf_t \rmd x.
\eea
A substitution into (\ref{molt_mu}), H\"older's inequality and a priori estimate (\ref{apr}) yield
\bea
&&\fl\frac{\kappa}{2}\frac{\rmd }{\rmd t }\|\nabla\vf\|^2 + \b \|\vf_t\|^2 + \gamma\|\nabla \mu\|^2 
\\
&&\leq
C\left[(1+ \|\bv_2\|_{\infty})\|\vf\|_{H^1} + \|\bv\|_4\right]\|\vf_t\| 
+ C[ \|\bv\|_4 + \|\bv_2\|_{\infty} \|\nabla\vf\| ]\|\mu\|.
\eea
Young's inequality, (\ref{L4}) and (\ref{infty}) lead to the estimate
\begin{eqnarray}\label{final_phi}
&&\fl\frac{\kappa}{2}\frac{\rmd }{\rmd t }\|\nabla\vf\|^2+ \frac{7}{8}\beta \|\vf_t\|^2 + \frac{7}{8}\gamma\|\nabla \mu\|^2 \\
\nonumber
&&\leq 
C(1+\|\bv_2\|^2_{H^2}) \|\vf\|^2_{H^1} + C\|\bv\|^2 + \frac{\nu}{14} \|\nabla\bv\|^2 + \eta_1 \|\mu\|^2,
\end{eqnarray}
where $\eta_1$ is a suitable positive constant.

By multiplying (\ref{diff_mu}) by $\mu$ and accounting for (\ref{apr}), we infer that
$$
\|\mu\|^2 \leq
 C\|\nabla\mu\|\|\nabla\vf\| + C[\|\vf_t\|+\|\bv_1\|_{H^2} \|\vf\|_{H^1}+ \|\bv\|_{H^1}+\|\vf\|_{H^1}  
 ]\|\mu\|.
$$
Hence
\begin{equation}
\label{mu_final}
\|\mu\|^2\leq
	 C[\|\nabla\mu\|^2 + \|\vf_t\|^2 +\|\bv_1\|_{H^2}^2 \|\vf\|_{H^1}^2+ \|\bv\|_{H^1}^2+\|\vf\|_{H^1}^2    ].
\end{equation}
By multiplying (\ref{diff-v}) by $\bv$ we obtain
\begin{equation}\label{mult-v}
\frac12 \frac{\rmd }{\rmd t }\|\bv\|^2 + \nu\|\nabla\bv\|^2 
\leq
I_1 + I_2 +I_3 + I_4,
\end{equation}
where
\bea
&&\fl I_1 = \kappa \int_\W (\nabla\vf_1 \otimes \nabla\vf + \nabla\vf \otimes \nabla\vf_2): \nabla\bv \rmd x
\\
&&\fl I_2 =  \l \int_\W (\vf\vf_{1t}\bv_1 + \vf_2\vf_t\bv_1 + \vf_2 \vf_{2t}\bv)\cdot \bv \rmd x
\\
&&\fl I_3 =  \l \int_\W [\vf(\bv_1\cdot \nabla\vf_{1})\bv_1 
+ \vf_2(\bv_1\cdot\nabla \vf_1)\bv + \vf_2(\bv\cdot\nabla\vf_1 + \bv_2 \cdot \nabla \vf)\bv_2]\cdot \bv  \rmd x\\
&&\fl I_4 = -\int_\W [(\nabla \bv) \bv_1 + (\nabla\bv_2) \bv] \cdot \bv \rmd x.
\eea
H\"older's, Young's inequalities, (\ref{Sob1}), (\ref{L4}) and a priori estimate (\ref{ap2}) allow us to estimate $I_1$, namely
\begin{eqnarray}\label{I2}
\fl I_1 \leq \kappa (\|\nabla\vf_1\|_4 + \|\nabla\vf_2\|_4) \|\nabla\vf\|_4 \|\nabla\bv\|
\leq
\frac{\nu}{14}  \|\nabla\bv\|^2 + \eta_2 \|\vf \|^2_{H^2} + C\|\nabla\vf\|^2.
\end{eqnarray}
Similarly, we obtain
\begin{eqnarray}\nonumber
&&\fl	I_2 \leq
	\l[\|\vf\|_{H^1} \|\vf_{1t}\| \|\bv_1\|_{H^1} + \|\vf_2\|_{H^1} \|\vf_{t}\| \|\bv_1\|_{H^1}
 + \|\vf_2\|_{H^1} \|\vf_{2t}\| \|\bv\|_4]\|\bv\|_4
	\\
	\label{I3}
	&&\leq 
	\frac{\nu}{14} \|\nabla\bv\|^2 + \frac{\beta}{8} \|\vf_t\|^2 + C(\|\bv\|^2 + \|\vf\|^2_{H^1})
	\end{eqnarray}
and
\begin{eqnarray}
	\nonumber
	&&\fl I_3 \leq
	\l [\|\vf\|_{H^1} \|\bv_1\|_{H^1}^2 \|\vf_1\|_{H^2} 
	 + (\|\bv_1\|_{H^1} +
	\|\bv_2\|_{H^1})\|\vf_2\|_{H^1}\|\vf_1\|_{H^2} \|\bv\|_4\\
	 &&+ 
	\|\vf_2\|_{H^1}\|\bv_2\|^2_{H^1}\|\nabla\vf\|_4]\|\bv\|_4
	\nonumber\\
	\label{I4}
	&&\leq 
	\frac{\nu}{14} \|\nabla\bv\|^2 + \eta_2\|\vf\|_{H^2}^2 + C(\|\vf\|_{H^1}^2 + \|\bv\|^2).
\end{eqnarray}
Finally, last integral can be controlled as
\begin{eqnarray}\label{I1}
\fl I_4 \leq \|\nabla \bv\| \|\bv_1\|_4 \|\bv\|_4 + \|\nabla\bv_2\| \|\bv\|_4^2
\leq
\frac{\nu}{14} \|\nabla \bv\|^2 +  C(\|\bv_1\|^2_{H^1} + \|\bv_2\|^2_{H^2}) \|\bv\|^2.
\end{eqnarray}

From (\ref{diff_mu}) it follows that
\bea
&&\fl\|\Delta\vf\| \leq C \Big[\|\mu\| + \|\vf\|_{H^1}\|\vf_1^2 +\vf_1\vf_2 + \vf_2^2\|_{H^1} + \|\vf\| + \|\vf\|_{H^1}\|\bv_1\|^2_{H^1} 
\\
 &&+ \|\vf_2\|_{H^1}\|\bv_1+\bv_2\|_{H^1}\|\bv\|_4+ \|\vf_t\| + \|\bv_1\|_{H^2} \|\nabla\vf\|
 +\|\nabla\vf_2\|_4 \|\bv\|_4 
\Big]
\\
&&\leq 
C (\|\mu\| + \|\vf\|_{H^1} + \|\vf_t\| + \|\bv\|_4) + C\|\bv_2\|_{H^2} \|\nabla\vf\|.
\eea
As a consequence, by means of (\ref{mu_final}), we have
\begin{eqnarray}
\nonumber
&&\fl\|\vf\|_{H^2}^2 \leq C(\|\vf\|^2 + \|\Delta\vf\|^2)\\
&&\leq C\left[ \|\nabla\mu\|^2  + (\|\bv_1\|_{H^2}^2 + \|\bv_2\|_{H^2}^2  +1 ) \|\vf\|^2_{H^1}\right.
\label{laplac}
\left. + \|\vf_t\|^2 + \|\bv\|^2_{H^1} \right].
\end{eqnarray}
Adding inequalities (\ref{diff_Sphi}), (\ref{final_phi}) and (\ref{mult-v}), accounting for (\ref{mu_final}), (\ref{laplac}) and choosing $\eta_1, \eta_2$ small enough, we prove the estimate
\bea
\fl\frac12\frac{\rmd }{\rmd t } [\|\vf\|^2 + \kappa \|\nabla\vf\|^2 +\|\bv\|^2]+ \frac{\b}{2} \|\vf_t\|^2 + \frac{\gamma}{2} \|\nabla \mu\|^2+ \frac{\nu}{2}\|\nabla\bv\|^2 
\leq  h(t)(\|\vf\|^2_{H^1}+\|\bv\|^2)
\eea
where 
$$
h(t) = C(\|\bv_1\|_{H^2}^2 + \|\bv_2\|_{H^2}^2 +1)
$$
is a $L^1-$function of time.
Thus, Gronwall's inequality proves 
$$
\vf=0, \qquad\qquad \bv={\bf 0}.
$$
Accordingly, from (\ref{diff_mu}) it follows that $\mu=0$ and we reach the conclusion.
\hfill$\square$

\bigskip

\noindent{\bf Acknowledgments.} The authors have been partially supported
by G.N.F.M. - I.N.D.A.M. through the project for young researchers
``Mathematical models for phase transitions in special materials".

\end{document}